\documentclass[journal,12pt, draftclsnofoot, onecolumn]{IEEEtran}
\usepackage{subcaption}
\usepackage[pdftex]{graphicx}
\usepackage{cite}
\usepackage{upgreek}
\usepackage{cite}
\usepackage{xcolor,soul,framed} 
\colorlet{shadecolor}{yellow}
\usepackage{url}
\usepackage{array}
\usepackage{caption}
\usepackage{balance}
\colorlet{shadecolor}{yellow}
\usepackage{color,soul}
\usepackage[cmex10]{amsmath}
\usepackage{algorithmic}
\usepackage{mdwmath}
\usepackage{mdwtab}
\usepackage{eqparbox}
\usepackage{array}
\usepackage{fixltx2e}
\usepackage{dblfloatfix}
\usepackage[mathscr]{euscript}
\usepackage{amsfonts}
\usepackage{amsmath}
\usepackage{commath}
\usepackage{upgreek}
\usepackage[long]{optidef}
\usepackage{bm}
\usepackage{algorithm}

\usepackage[open,openlevel=2]{bookmark}
\usepackage{cleveref}
\usepackage{textgreek}
\usepackage{booktabs,chemformula}
\usepackage{float}
\usepackage{amssymb}
\usepackage{bm}
\usepackage[thinlines]{easytable}
\usepackage{comment}
\newcolumntype{P}[1]{>{\centering\arraybackslash}p{#1}}

\begin{document}
\DeclareGraphicsExtensions{.png}
\title{Study of BSM Inter-Packet Gap Tails in C-V2X Networks}
\author{Abdurrahman~Fouda,~\IEEEmembership{Student Member,~IEEE,}
        Randall~Berry,~\IEEEmembership{Fellow Member,~IEEE,}\\
        and Ivan~Vukovic~\IEEEmembership{Senior Member,~IEEE}
        
    \thanks{This research is supported in part by Ford Motor Company. An earlier conference version of this article was presented in part in IEEE Vehicular Networking Conference (VNC) 2021~\cite{vnc}.}
    \thanks{A. Fouda and R. Berry are with the Department of Electrical and Computer Engineering, Northwestern University, Evanston, IL, 60208 USA (email: abdurrahman.fouda@u.northwestern.edu, rberry@northwestern.edu).} 
    \thanks{I. Vukovic is with Ford Motor Company, Dearbon, MI, 48124 USA (email: ivukovi6@ford.com).}}
\markboth{}%
{Shell \MakeLowercase{\textit{et al.}}: Bare Demo of IEEEtran.cls for IEEE Journals}

\maketitle
\vspace{-0.5in}
\begin{abstract}

Cellular vehicle-to-everything (C-V2X) enables safety-critical connected vehicular services by exchanging basic safety messages (BSMs) among nearby vehicular users (VUEs). Timely transmission of BSMs is crucial to avoid stale information at VUEs. However, successive packet losses can lead to large inter-packet gaps (IPGs), reducing the BSMs’ reliability. This paper investigates the tail behavior of IPG and information age (IA) distributions in C-V2X mode 4, a decentralized resource allocation method based on semi-persistent scheduling (SPS). Specifically, we study the improvements and trade-offs introduced by the SAE-specified concept of one-shot transmissions to decrease the number of successive BSM losses at destination VUEs. The study employs a high-fidelity system-level simulator that closely follows the SPS process of C-V2X mode 4 to evaluate the performance of interleaved one-shot SPS transmissions. The numerical results demonstrate significant improvement in the IPG and IA tail distributions in various simulation scenarios. Additionally, we propose an accurate analytical model to characterize the IPG tail behavior of C-V2X BSM transmissions. The proposed model is validated by comparing its results with those obtained using the system-level simulations. Our validation demonstrates that the proposed model generates analytical results that coincide with the asymptotic slopes of IPG distribution in different BSM transmission modes.

\end{abstract}

\IEEEpeerreviewmaketitle

\section{Introduction}\label{sec_intro}
\IEEEPARstart{V}{ehicle}-to-everything (V2X) communications have attracted a lot of academic and industrial attention over the last decade as a way to improve vehicle's reliability and efficiency. Recently, cellular V2X (C-V2X) has emerged as a leading approach for delivering V2X services. There has been a growing interest from several automakers and regulators to use C-V2X~\cite{toyota,Renault,asilomar}. In this regard, the 3rd Generation Partnership Project (3GPP) has introduced support of C-V2X services (also referred to as LTE-V2X) in Releases 14 and 15~\cite{3gpp37885}. Further enhancements to support advanced new radio V2X (NR-V2X) operation scenarios are introduced in Releases 16 and 17~\cite{3gpp38885},~\cite{nrpos}. It is expected that the support of advanced use cases and new wireless technologies for V2X will continue in Release 18 and beyond~\cite{5gaa_slr}. Currently, 3GPP has introduced transmission modes 3 and 4 for C-V2X systems extending the LTE device-to-device (D2D) transmission modes 1 and 2~\cite{3gpp36213}. In mode 3, base stations (eNBs) are responsible for the radio resource management (RRM) process. Our focus in this paper is on mode 4, in which, data and control packets are exchanged directly between vehicular users (VUEs) without assistance from eNBs. In particular, VUEs use a distributed radio resource allocation scheme,  namely, semi-persistent scheduling (SPS), to autonomously select their radio resources~\cite{3gpp36213},~\cite{3gpp36321}. 


C-V2X safety-related services (e.g., collision avoidance) are instrumental for improving road safety and traffic efficiency. Given that these applications are time-critical, the reliability and latency of communications are key metrics. Two common metrics for evaluating the latency of V2X communications are information age (IA) and the inter-packet gap (IPG)~\cite{5gaa_PerforTest}. It has been observed that the IPG (and IA) distribution in C-V2X systems can exhibit a long tail because the persistent nature of SPS can result in successive packet losses at the receiver VUEs~\cite{toyota,bspots}. The Society of Automotive Engineers (SAE) C-V2X technical Committee has specified a solution to remedy this by using \textit{one-shot transmissions} to decrease the probability of persistent packet collisions. However, introducing another degree of randomness in the SPS scheduler such as one-shot transmissions may involve a trade-off with the average number of successfully received packets at destination VUEs. In this paper, we seek to study the impacts of using one-shot transmissions in C-V2X systems to improve the IPG and IA tail behavior and the trade-offs with average packet reception rate (PRR) through accurate discrete-event system simulations and analytical modeling.     

The SAE has introduced basic safety messages (BSMs) for vehicular safety communications in V2X networks~\cite{J2735}. BSMs enable VUEs to share relative speed, position, and other mobility information with nearby VUEs. SAE is in the process of standardizing BSM transmissions using C-V2X for vehicle-to-vehicle (V2V) scenarios~\cite{J3161}. C-V2X transmission mode 4 is a natural selection for V2V communication because VUEs do not need to rely on the availability of cellular coverage for successful communication~\cite{tm4anal}. Further, VUEs can leverage the decentralized nature of mode 4 to avoid interaction with eNBs and decrease the transmission latency, control overhead, and frequent handovers. C-V2X mode 4 defines V2V communication which is referred to in 3GPP terminology as sidelink (PC5 being the interface name) and is being specified in~\cite{J3161}. The sidelink performance of BSM transmission using mode 4 has been studied extensively in the literature to evaluate the reliability, in terms of packet reception ratio (PRR), and latency, in terms of IPG and/or IA, via: new collaboration methods between VUEs~\cite{piggyback, keep}, SPS parameter tunning~\cite{AnalVNC, TM4Config}, different congestion control mechanisms~\cite{Apc, spatio}, enhanced SPS schemes~\cite{esps, AugRA}, analytical modeling to calculate the packet collision probability~\cite{xuwork}, and by comparing DSRC (another vehicular communication protocol) with C-V2X via simulation campaigns~\cite{toyota, Renault, asilomar, HIL}. Further, power optimization-based frameworks~\cite{ProbAoI1, ProbAoI2}, and AI-aware RRM schemes~\cite{RRM} have also been investigated to improve the IA tail (i.e., the worst-case scenarios) of V2X transmission mode 3. In addition, the authors in~\cite{bspots} proposed an alternative method to decrease the probability of losing consecutive BSMs without impacting the transmission reliability by limiting the maximum duration, at which, a VUE keeps the same resources for BSM transmission.

However, none of the prior studies have discussed the tail improvement of the IPG distribution at destination VUEs using one-shot transmissions based on SAE draft specification~\cite{J3161}. Moreover, the results in~\cite{bspots} show a related tail improvement of the \textit{wireless blind spot duration} which measures the time a vehicle stays in a given choice of colliding resources. This is related to the IPG length but does not account for other causes of packet losses or the fact that vehicles can reselect into another collision, further extending the IPG. To the best of the authors' knowledge, none of these contributions have proposed an analytical model to characterize the IPG tail of C-V2X transmission mode 4. Furthermore, new standardized signaling is needed to incorporate the proposed VUE cooperation and scheduling methods considered in some prior work, which is not the case for the one-shot transmissions considered here.

This paper considers intertwining the concept of one-shot transmissions as specified in~\cite{J3161} with the sensing-based SPS to improve the IPG and IA tails of BSM transmissions using C-V2X mode 4. By randomly skipping the SPS reserved radio resources, the one-shot transmissions mechanism seeks to decrease the probability of persistent BSM collisions at destination VUEs. Our contributions are summarized as follows:
\begin{itemize}
    \item Extensive simulation campaigns are conducted using a C++ simulator that closely follows the SPS process of C-V2X transmission mode 4 specified in~\cite{3gpp36213} and the one-shot transmissions from~\cite{J3161} (formally known as event based transmissions). The performance of the interleaved one-shot SPS transmissions is evaluated at different bandwidth configurations, vehicle densities, and V2V distances in terms of IA, IPG, PRR, channel busy ratio (CBR), and maximum inter-transmit time (ITT).
    \item An analytical model is developed to approximate the tails of the IPG and IA complementary cumulative distribution functions (CCDF) of BSM updates with and without one-shot transmissions. The proposed model is validated by comparing its results with those obtained by the C-V2X system-level simulations.
\end{itemize}

The numerical results show that the one-shot transmissions can significantly improve the IA and IPG CCDF tails compared to the SPS implementation solely based on~\cite{3gpp36213}. We also demonstrate that the improvement in the 99.9th-percentile of the IA and IPG is robust against different simulation scenarios. Finally, our robust validation of the proposed analytical model shows that it closely coincides with the asymptotic slopes of the IPG CCDFs of different simulation scenarios. The findings in this paper are as follows:
\begin{itemize}
    \item The simulation results reveal that the IPG and IA tail improvements are influenced by the vehicle density, Tx-Rx separation, and bandwidth configuration. In particular, the best tail improvement can be achieved at the low vehicle densities compared with the high vehicle densities.
    \item  Moreover, we show that the closer the Tx-Rx pairs are, the better the tail improvement is. The numerical results also demonstrate that the expected degradation in PRR performance (due to the interrupted persistency of SPS) is small compared to the significant improvements in the IA and IPG tails.
    \item The proposed probabilistic model reveals that the IPG long-tail behavior is driven by the received interference at the receiver VUEs. Specifically, we show that interference is the dominant reason (among other reasons such as half-duplex (HD) transmissions and low levels of the received signal power below the thermal noise) for the successive packet losses at destination VUEs.
    \item The analysis also shows that it is sufficient to assume that the resource reselection is the dominant reason for ending the long IPGs at destination VUEs when congestion control is not in operation. In contrast, it is necessary to consider another effect, which we refer to as \textit{slippage}, when congestion control is used. \textit{Slippage} represents the transmission scenarios, at which, the transmissions of two colliding vehicles go out of phase when BSM generation rate is not an integer multiple of 100 milliseconds (ms).
\end{itemize}

The rest of this paper is structured as follows. The SPS baseline scheduling scheme and one-shot transmissions are presented in Section~\ref{sec_tm4}. In Section~\ref{sec_results}, we show the performance gains of using the interleaved one-shot SPS transmissions for BSM updates. Section~\ref{sec_analmodel} discusses the proposed analytical model to estimate the IPG CCDF tail behavior of BSM transmission in C-V2X systems. Finally, concluding remarks are given in Section~\ref{sec_conc}.

\section{Transmission Mode 4 in C-V2X Networks}\label{sec_tm4}

We consider a V2V broadcast scenario for BSM transmissions in a C-V2X system. Specifically, a set of VUEs utilizes C-V2X transmission mode 4 to periodically broadcast heartbeat BSMs to their neighbor vehicles in a HD way. In doing so, VUEs transmit their data packets using the physical sidelink shared channel (PSSCH). The control packets are transmitted via the physical sidelink control channel (PSCCH) using the sidelink control information (SCI) format 1~\cite{3gpp36212}. SCI carries the necessary information to decode the data packets transmitted via PSSCH, such as the modulation and coding scheme (MCS), frequency resource locations of the initial (re)transmission, packet priority, and retransmission index. SCI also contains the necessary information to assist VUEs in the SPS resource (re)selection process to reduce packet collisions (e.g., the value of the resource reselection counter~\cite{3gpp36212}) as will be discussed later.


In C-V2X, the physical channel is divided into sub-frames in the time domain and sub-channels in the frequency domain. The sub-frame width is 1 ms (i.e., a transmission time interval (TTI)), representing the time granularity for message scheduling in C-V2X transmission mode 4. The minimum allocation unit in the frequency domain for an LTE user is a physical resource block (PRB). A PRB spans 180 kHz in the frequency domain, 0.5 ms in the time domain, and contains 12 subcarriers separated by 15 kHz each. C-V2X defines a sub-channel as the minimum allocation unit to a VUE in the frequency domain. A sub-channel occupies a configurable number of PRBs, and the number of sub-channels per sub-frame is determined based on the operating bandwidth. 

In this paper, we refer to a sub-channel as a virtual resource block (VRB) and use the configuration of 10 PRBs per VRB, where, e.g., 20 MHz bandwidth consists of 10 VRBs~\cite{3gpp36331}. In C-V2X, data packets (e.g., BSMs) are transmitted in transport blocks (TBs). A TB consists of contiguous PRBs per sub-frame. The number of PRBs per TB depends on the size of the transmitted data packets and the MCS. As mentioned earlier, data packets are sent along with SCI, which occupies two contiguous PRBs per sub-frame. It is worth noting that the BSM TBs and SCI PRBs can be either adjacent or non-adjacent and are only required to be in the same sub-frame~\cite{3gpp36213}. For simplicity, we assume that BSM TBs and the corresponding SCI are always adjacent. Further, we use a fixed packet payload size for BSM and SCI transmission, which occupies two contiguous VRBs per sub-frame~\cite{J2945}. The first two PRBs are reserved for SCI, and the remaining PRBs are reserved for data packets.

\subsection{Semi-persistent Scheduling}\label{sec_sub_sps}
In C-V2X transmission mode 4, VUEs utilize a sensing-based SPS scheduling scheme to autonomously allocate the radio resources without assistance from the cellular infrastructure. VUEs randomly select the required VRBs for BSM transmission from a candidate list of VRBs. The candidate list is defined using a pre-configured resource pool,  namely, the selection window. The selection window size in the time domain is given by $[n+T_{1},n+T_{2}]$. Here, $n$ denotes the BSM generation time, $T_{1}\le{4}$, and $T_{2}$ is determined based on the packet delay budget (PDB) $\xi$ where $T_{2}=\mathrm{max}(\xi-10,\,20\,\mathrm{ms})$~\cite{J3161}. Here, $\xi$ is defined as the maximum allowed latency between the BSM generation time and the actual transmission time. It can be determined using Table~\ref{tab_spsparams} based on $\tau$ which denotes the BSM generation rate (i.e., the total number of transmitted packets per second). It is worth mentioning that the BSM generation time refers to the BSM arrival time from the application layer to the physical layer, at which, the BSM becomes ready to be transmitted to neighboring VUEs.

To facilitate SPS scheduling, SCIs are considered only if something is received and, if so, the average received reference signal resource power (RSRP) is used as a metric to exclude VRBs (from the selection window) whose RSRP is above a given threshold. Each VUE also excludes VRBs which are used by other VUEs from the selection window. These VRBs are determined based on the received SCI in the last 1000 sub-frames. The total number of available VRBs for reselection should represent at least 20\% of all resources in the resource selection pool. If not, a VUE keeps increasing the RSRP threshold by 3 dB iteratively until the 20\% target is met. The received signal strength indicator (RSSI) of the PSSCH VRBs is then used as a sorting metric to determine the best 20\% VRBs that experience the lowest received RSSI. A VUE selects two contiguous VRBs of these resources randomly for the BSM transmission. 


Once a VUE selects a new set of VRBs, it keeps \textit{reusing} them persistently for the next consecutive $C_{\mathrm{s}}$ BSM transmissions. Note that \textit{reusing} VRBs means using the same VRBs in the same sub-frame (which is determined based on the BSM generation time $n$ and the PDB length (see Table~\ref{tab_spsparams})) each SPS selection window. Initially, once a BSM becomes ready to be transmitted at a given VUE, the VUE establishes the selection window and chooses the same sub-frame within that window for the next consecutive $C_{\mathrm{s}}$ BSM transmissions. Here, $C_{\mathrm{s}}$ is defined as the resource reselection counter (also referred to as the SPS interval) and is chosen uniformly at random between $\left[\alpha_\mathrm{s},\,\beta_\mathrm{s}\right]$, where $\alpha_\mathrm{s}$ and $\beta_\mathrm{s}$ are fixed integers with $0<\alpha_\mathrm{s}<\beta_\mathrm{s}$. The reselection counter is decremented by one after each BSM transmission. When $C_{\mathrm{s}}$ reaches zero, a VUE reselects a new set of VRBs with a reselection probability $p_\mathrm{r}=1-p_{\mathrm{p}}$. Here, $p_{\mathrm{p}}$ is the probability to keep the current VRBs for the next BSM transmission after $C_{\mathrm{s}}$ transmissions, where $p_{\mathrm{p}}\in[0,0.8]$ with a step of 0.2~\cite{J3161}. Once new VRBs are chosen or the current VRBs are kept, a new SPS interval then begins. 

\begin{table}
\renewcommand{\arraystretch}{.75}
  \centering
  \caption{Sensing-based SPS parameters~\cite{3gpp36213} ,~\cite{3gpp36321}.}\label{tab_spsparams}
        \begin{tabular}{ P{.7cm} P{2.5cm} P{2.5cm} }
        \hline
        \textbf{$\boldsymbol{\tau}$}  & \textbf{PDB $(\xi)$} & \textbf{$C_{\mathrm{s}}$ interval}\\ \hline
        10 & 100 ms & {$[5,\,15]$}\\ \hline
        20  & 50 ms & [10, 30]\\ \hline
        50   & 20 ms & [25, 75]\\ \hline
        \end{tabular}
        \vspace{-.25in}
\end{table}

\subsection{One-shot transmissions}\label{sec_sub_1shot}
This section discusses how one-shot transmissions can be intertwined with SPS transmissions to improve the BSM tail behavior of C-V2X transmission mode 4. Generally, one-shot transmissions rely on the idea of adding more randomness to the resource reselection process to avoid long IPGs (i.e., persistent packet collisions). Let $C_{\mathrm{o}}$ denote the one-shot resource reselection counter that is chosen uniformly at random between $[\alpha_\mathrm{o},\,\beta_\mathrm{o}]$, where $\alpha_\mathrm{o}$ and $\beta_\mathrm{o}$ are fixed integers with $0<\alpha_\mathrm{o}<\beta_\mathrm{o}$. When one-shot transmissions are used, a VUE decrements both $C_{\mathrm{s}}$ and $C_{\mathrm{o}}$ by one every packet transmission. Next, we discuss how one-shot transmissions are implemented by considering the possible relations between $C_{\mathrm{s}}$ and $C_{\mathrm{o}}$.   

\begin{figure}
  \begin{center}
  \includegraphics[width=8cm,height=8cm,,keepaspectratio]{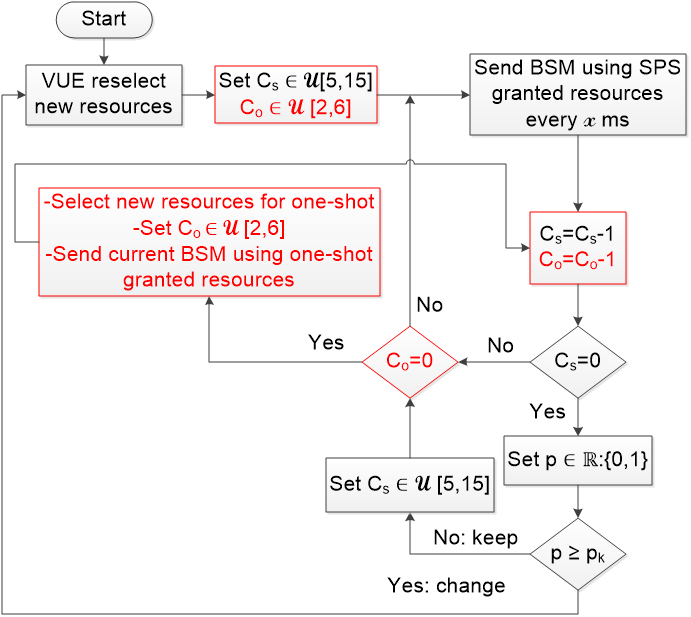}
  \caption{Flow chart of the interleaved one-shot SPS.}\label{fig_flowchart}
  \vspace{-.25in}
  \end{center}
\end{figure}

First, when $C_{\mathrm{s}}$ reaches zero, while $C_{\mathrm{o}}>0$, the VUE uses $p_{\mathrm{r}}$ to determine whether a new set of VRBs will be reselected or not. If a new set of VRBs is reselected, the VUE resets both counters and starts the process again. Otherwise, the VUE resets only the SPS counter $C_{\mathrm{s}}$ and decrements $C_{\mathrm{o}}$ by one. Second, when $C_{\mathrm{o}}$ reaches zero, a new set of resources is reselected and used only for the current transmission opportunity. The one-shot VRBs are selected using the same sensing-based reselection process discussed in Section~\ref{sec_sub_sps}. The VUE then resets $C_{\mathrm{o}}$ and uses the regular SPS-granted VRBs for the next BSM transmission opportunity. Finally, when both $C_{\mathrm{s}}$ and $C_{\mathrm{o}}$ reach zero simultaneously, the VUE rests both counters and uses $p_{\mathrm{r}}$ to determine whether a new set of VRBs will be reselected. If VUE decides to keep the old VRBs, it reselects a new set of one-shot VRBs and uses them for the current transmission opportunity. The old set of SPS-granted VRBs is then used for BSM transmission in the next transmission opportunity. If a new set of VRBs is reselected, VUE keeps using it for BSM transmission until either of $C_{\mathrm{s}}$, $C_{\mathrm{o}}$, or both of them expire and then repeats the above steps. The interleaved one-shot SPS is summarized in Fig.~\ref{fig_flowchart}, in which, the black steps represent the legacy SPS and the red steps denote the additional steps for adding one-shot transmissions. 

\subsection{SAE Congestion Control}\label{sec_sub_cc}
In C-V2X networks, the packet loss rate increases when the number of competing vehicles over the same wireless resources increases~\cite{capacity}. Hence, congestion control is vital to provide VUEs with adequate levels of successful packet reception. A decentralized congestion control mechanism for the application layer (which we use in this paper) is introduced in SAE J3161/1 to decrease the number of packet collisions in highly congested scenarios~\cite{J2945},~\cite{J3161}. In this approach, each VUE adjusts its BSM generation rate based on the estimated number of nearby transmitting vehicles. Let $N^{v}_{\mathrm{c}}$ denotes the current number of unique neighbor vehicles in a given range of $r$ meters that the $v^{\mathrm{th}}$ VUE has detected at least once in the previous 1000 ms. Note that, a neighbor vehicle is determined to be unique if it has a unique ID in its BSM transmission. 

The average smoothed vehicle density at $v^{\mathrm{th}}$ VUE can be calculated at $k^{\mathrm{th}}$ sub-frame as: 
\begin{equation}
    N_{\mathrm{s}}^{v}(k)=\lambda\times{N^{v}_{\mathrm{c}}}+(1-\lambda)\times{N^{v}_{\mathrm{s}}(k-1)}.
\end{equation}
In that, $N_{\mathrm{s}}^{v}(k)$ is updated every 100 ms using the current number of detected unique neighbor vehicles that is measured over a sliding window of 1000 ms. Here, $\lambda$ is a weight factor between the current vehicle density and the previously estimated density. Now, let $I(k)$ represent the time interval between BSM packet generations in ms for the $v^{\mathrm{th}}$ VUE at the $k^{\mathrm{th}}$ sub-frame. SAE J3161/1 specifies that $I(k)$ be adapted based on the average estimated vehicle density as follows:   
\begin{equation}\label{eq_cc}
        I(k)=
        \begin{cases}
            100 & \,\,\,N_{\mathrm{s}}^{v}(k) < B,\\
            100\times\frac{{N_{\mathrm{s}}^{v}(k)}}{B} & \,\,\,B \leq{N_{\mathrm{s}}^{v}(k)}<\frac{I_{\mathrm{max}}}{100}\times{B},\\
            I_{\mathrm{max}} &\,\,\, \frac{I_{\mathrm{max}}}{100}\times{B}\leq{N_{\mathrm{s}}^{v}(k)},
        \end{cases}
\end{equation}
where $B$ and $I_{\mathrm{max}}$ denote the vehicle density coefficient and the maximum allowed BSM generation interval, respectively. In other words, the generation rate of BSMs decreases linearly when the estimated vehicle density exceeds $B$ until it reaches a minimum value of $1/I_{\mathrm{max}}$. Note that $I(k)$ is updated every 100 ms regardless of the BSM generation rate. Also, note that the above congestion control mechanism controls when BSMs are generated. Once generated, a BSM is then sent in the next sub-frame selected via the SPS or one-shot reselection.  When congestion control is activated a VUE will not transmit in the selected sub-frame unless a new BSM has been generated.

\section{Numerical Analysis}\label{sec_results}

\begin{table}
\renewcommand{\arraystretch}{0.75}
  \centering
  \caption{Simulation parameters.}\label{tab_simpara}
        \begin{tabular}{l l}
        \hline
            \textbf{Parameter}  & \textbf{Value}\\
        \hline
            Scenario & Layout: 5 km single-lane Highway, Density: 125, 400 and 800 VUE/km\\
        \hline
            Channel model  & Path loss: ITU-R UHF urban canyon~\cite{ITUpathloss}, In-band emission (IBE): 3GPP TR 36.885~\cite{3gpp37885}\\
            & Fast fading: Nakagami-$m$ distribution\\
        \hline
            Antenna settings & $n_t=1$, $n_r=2$, MRC receiver\\
        \hline
           One-shot settings & $C_{\mathrm{o}}\in[2,\,6]$ and $[5,\,15]$\\
        \hline 
           SPS settings & $T_\mathrm{1}=4$, $T_\mathrm{2}=90$, $\xi=100$ ms, $p_{\mathrm{p}}=0.8$, $C_{\mathrm{s}}\in{[5,\,15]}$, CBR threshold: -94 dBm\\
        \hline
            Resource pool settings & Carrier bandwidth: 10, 20 MHz at 5.9 GHz, Sub-channels per sub-frame: 5, 10\\
            & PRBs per sub-channel: 10\\ 
        \hline
            Power settings   & Tx power: 20 dBm, Noise figure: 6 dB, Thermal noise: -174 dBm / Hz\\ 
            & PSCCH power boost: 3dB \\
        \hline
            SINR thresholds   & PSSCH: 3 dB, PSCCH: 0 dB\\
        \hline
            BSM packet size & Payload size: 300 bytes, PSCCH PRBs: 2, Sub-channels/BSM: 2, PSSCH PRBs: 18.\\
        \hline
            Congestion control & $r,\,I_{\mathrm{max}},\,B,\,\lambda=100$ m, $600$ ms, $25$, $0.05$\\
        \hline
            Simulation time & 500 seconds\\
        \hline
        \end{tabular}
        \vspace{-.25in}
\end{table}
\begin{figure*}
\subfloat[Distance bin 200 m]{\includegraphics[width=.33\textwidth]{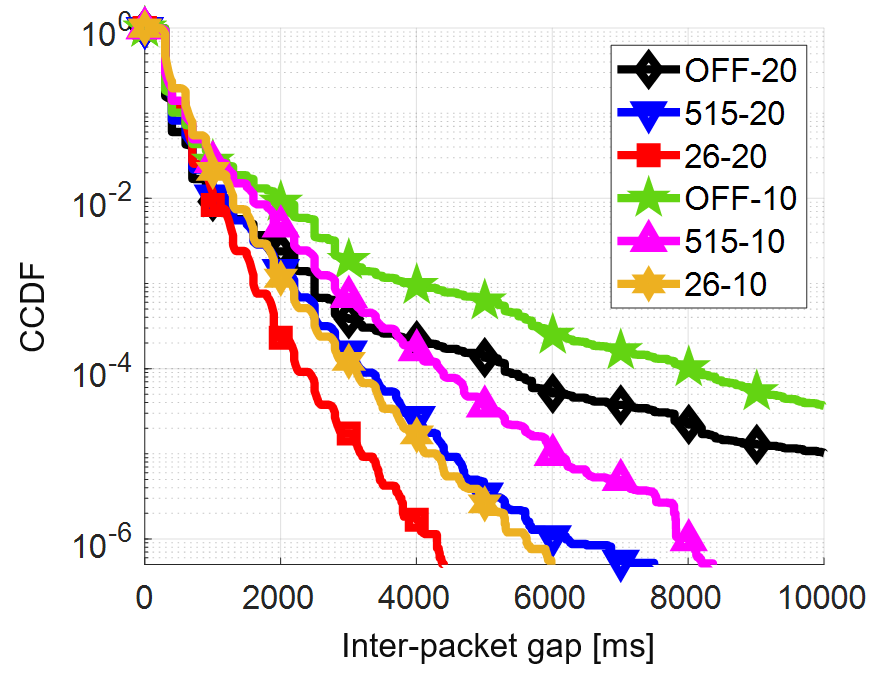}} 
\subfloat[Distance bin 300 m]{\includegraphics[width=.33\textwidth]{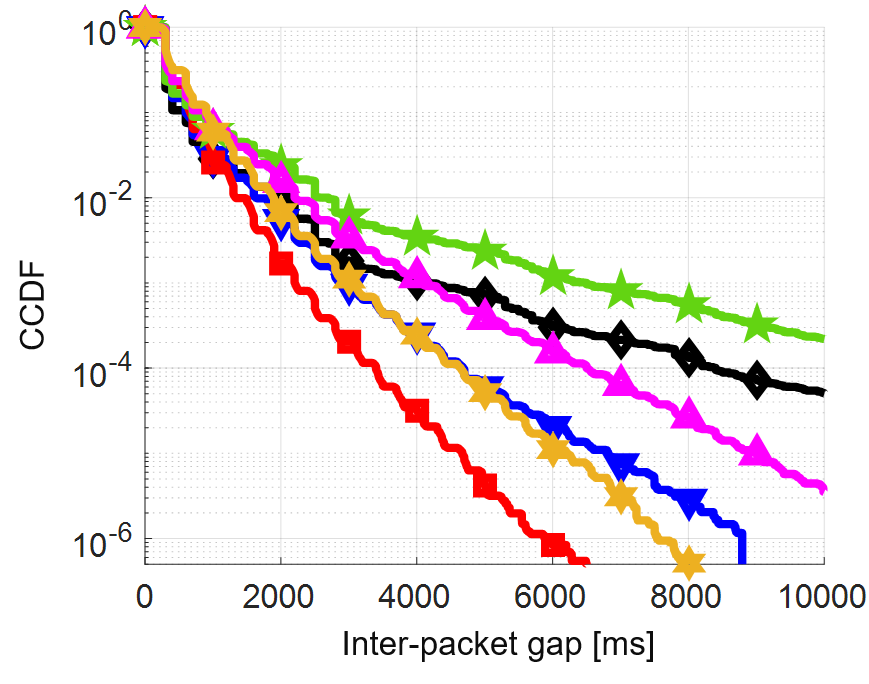}} 
\subfloat[Distance bin 400 m]{\includegraphics[width=.33\textwidth]{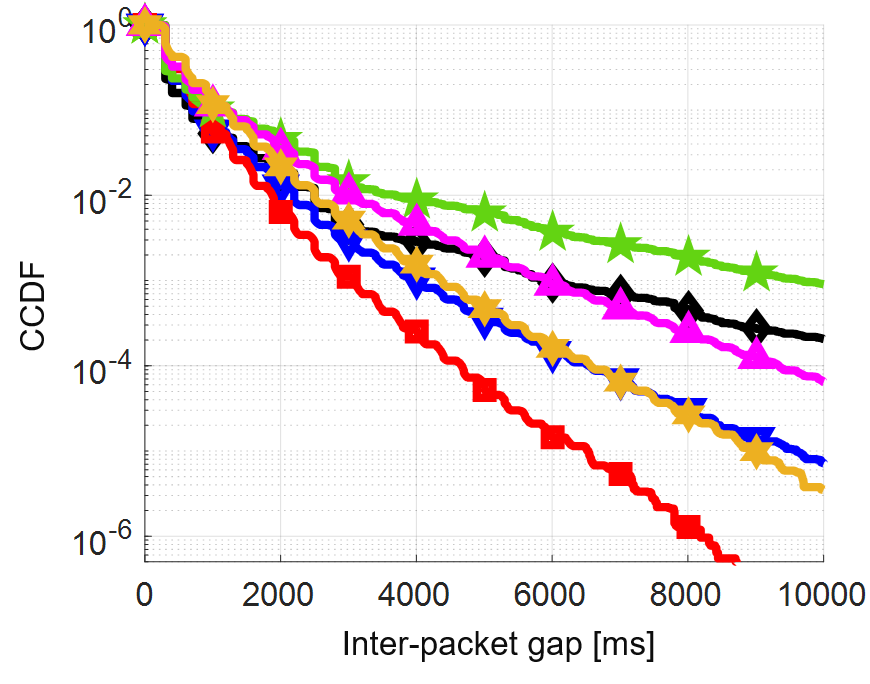}} 
\caption{CCDF of the inter-packet gap.}
\vspace{-.25in}
\label{fig_ipgccdf}
\end{figure*}
In this section, we illustrate the performance gains of using different one-shot configurations for BSM transmissions in C-V2X systems. We use a high-fidelity system-level C++ simulator that closely follows the SPS scheme for BSM transmission using C-V2X mode 4. Specifically, Monte Carlo simulations have been conducted to study the performance characteristics of different merit figures (e.g., IPG, IA, PRR, and CBR) for 10 MHz and 20 MHz bandwidths. In doing so, we use the ITU-R urban canyon path loss model in~\cite{ITUpathloss} with a Nakagami-$m$ distribution, which is given by:
\begin{equation}
    m(d)=
        \begin{cases}
            3 & \,\,\,d < 50,\\
            1.5 & \,\,\,50 \leq{d}<{150},\\
            1 &\,\,\, d\ge{150},
        \end{cases}
\end{equation}
where $m$ and $d$ denote the Nakagami fading parameter and the V2V distance in meters, respectively. Here, $\mathrm{Nakagami}(m,\,\Omega)$ random variates are generated using the $\mathrm{Gamma}(k,\,\theta)$ distribution as discussed in~\cite[Ch.~8]{SimulationModeling}, where $\Omega$ denotes the received non-faded power, $k=m $, and $\theta=\Omega/m$.

VUEs are regularly spaced over a single lane in a highway scenario following the simulation settings in Table~\ref{tab_simpara}. Each vehicle is equipped with $n_t$ transmit and $n_r$ receive antennas and performs maximal ratio combining (MRC) on the received BSM~\cite[Ch.~3]{D.Tse}.  We use the IBE models defined in\cite{3gpp37885} for 10 and 20 MHz bandwidths to account for the interference between different sub-channels in the same sub-frame. The link-level performance of the PSSCH and PSCCH is implemented using the  block error rate (BLER) versus SINR curves in~\cite{toyota} and~\cite{NIST}, respectively. A 3dB power boost is used to achieve an adequate performance of the PSCCH~\cite{3gpp36213}. We run the simulation scenarios for 500 seconds and collect the statistics only after the first 10 seconds, which is used as a simulation warm-up time to prevent any start-up bias. Numerical statistics are collected from vehicles located only in the middle third of the highway to minimize the boundary effects. In this work, we focus on investigating the advantage of using one-shot transmissions to decrease the number of consecutive BSM losses. Unless mentioned otherwise, we use a medium vehicle density of 400 VUE/km.

\subsection{Inter-packet Gap (IPG)}\label{sec_sub_ipg}
\begin{figure}
  \begin{center}
  \includegraphics[width=8cm,height=8cm,,keepaspectratio]{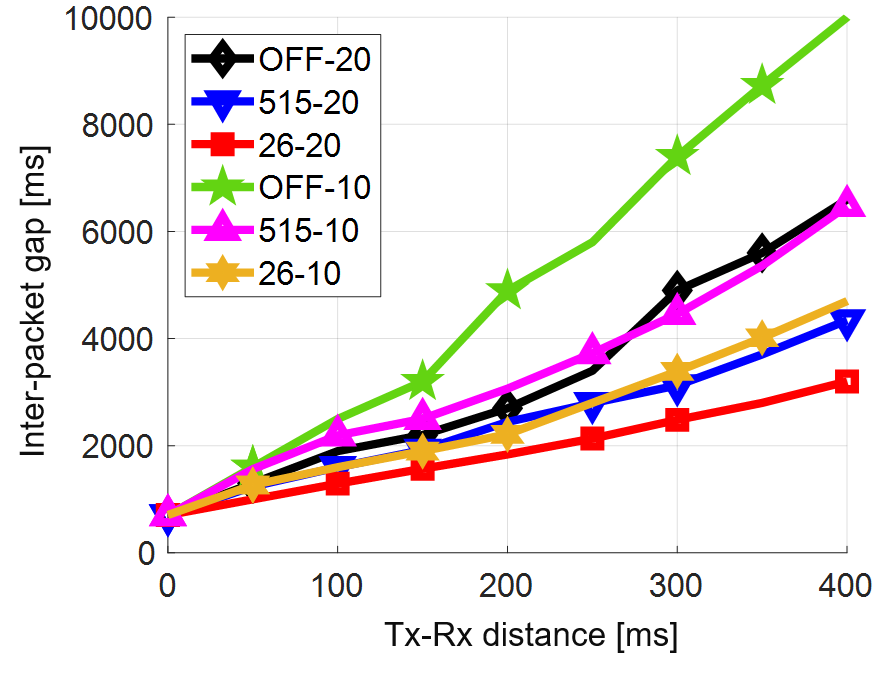}
  \caption{99.9th-percentile of inter-packet gap.}\label{fig_ipg999}
  \vspace{-.25in}
  \end{center}
\end{figure}

\begin{figure*}
\subfloat[Distance bin 200 m]{\includegraphics[width=.33\textwidth]{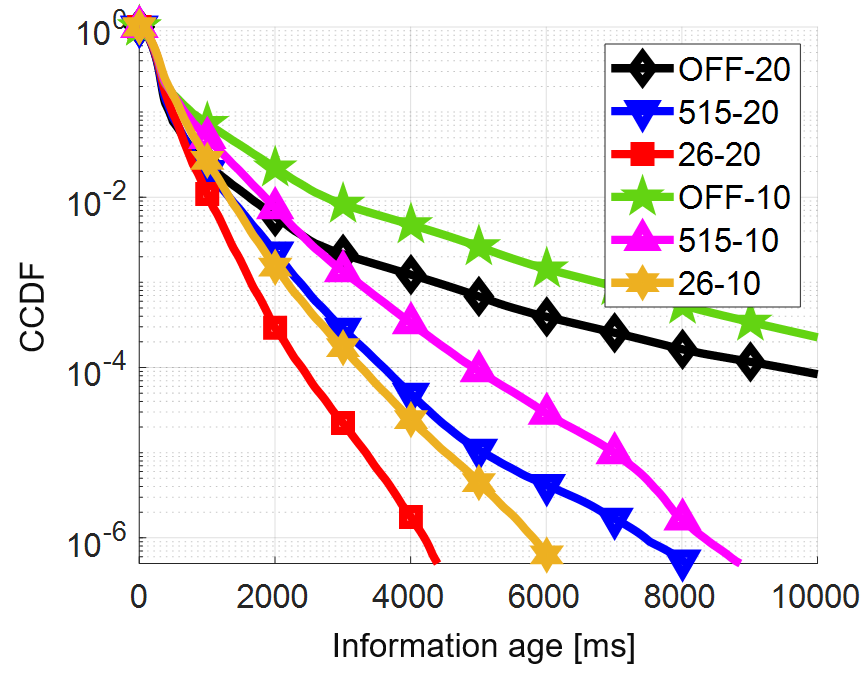}} 
\subfloat[Distance bin 300 m]{\includegraphics[width=.33\textwidth]{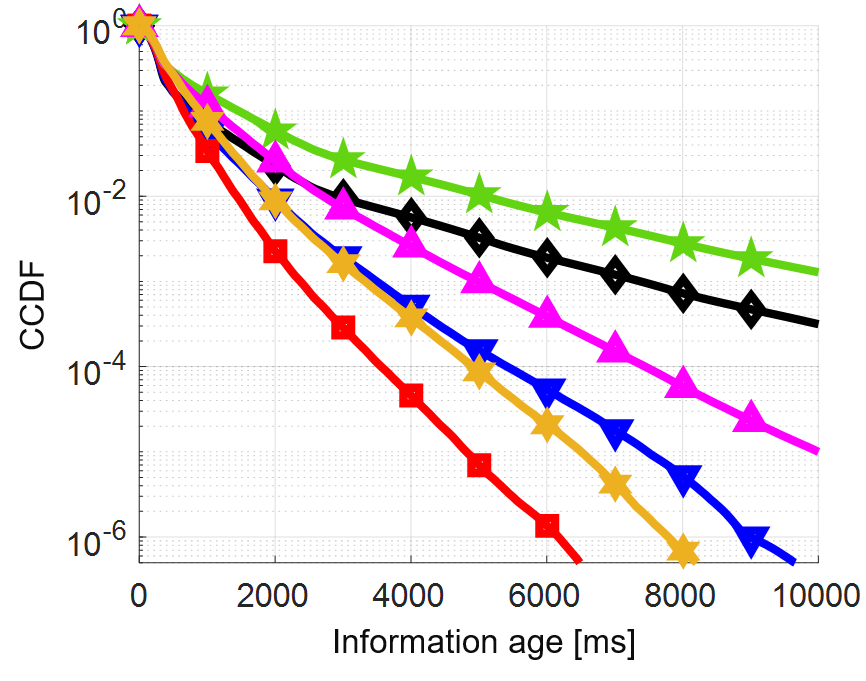}} 
\subfloat[Distance bin 400 m]{\includegraphics[width=.33\textwidth]{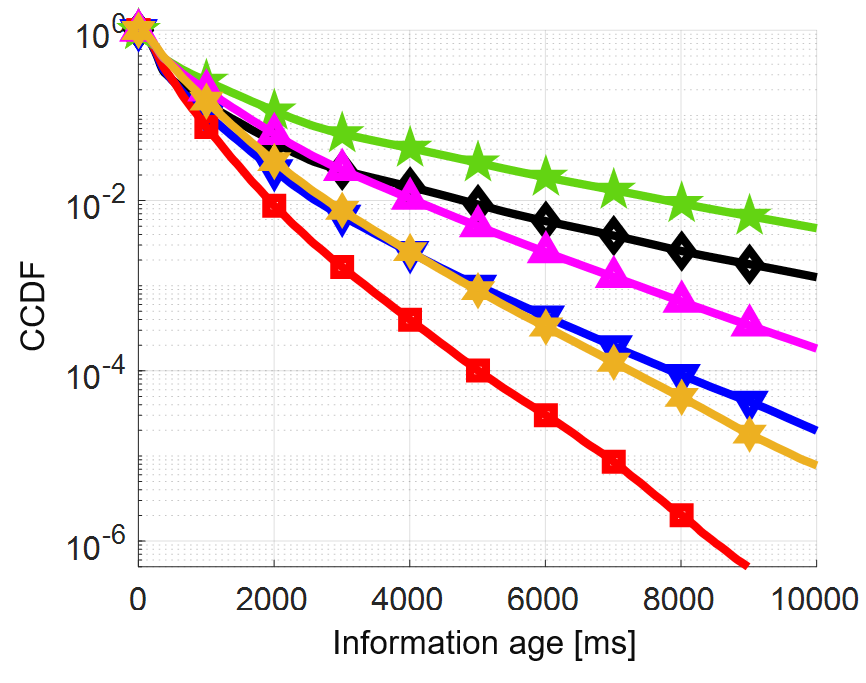}} 
\caption{CCDF of the information age.}
\label{fig_iaccdf}
\end{figure*}

In this section, we evaluate the potential gains of using one-shot transmissions to improve the IPG CCDF tails in C-V2X systems. For each pair of vehicles separated by distance $d$ meters, we measure IPG as the time elapsed between each successive pair of successfully received BSMs. The IPG CCDF $F(i)$ is then given by the fraction of those values that exceed $i$ milliseconds. Fig.~\ref{fig_ipgccdf} shows the IPG CCDF with and without one-shot transmissions. The statistics are collected for 10, 20 MHz bandwidths at distance bins 200, 300, and 400 m. A V2V distance bin of $d$ represents all Tx-Rx pairs that are separated by $\zeta$ meters, where $\zeta\in{[d-25,\,d+25]}$. We use two configurations for one-shot transmissions, in which, $C_{\mathrm{o}}$ can be chosen uniformly at random between $[2,\,6]$ or $[5,\,15]$. The simulation scenarios are denoted by $x$-$y$, where $x$ represents the $C_{\mathrm{o}}$ configuration and $y$ represents the operating bandwidth. As shown in Fig.~\ref{fig_ipgccdf}, with high probability, the IPG is $\approx$ 310 ms at all distance bins (i.e., vehicles successfully receive BSMs every 310 ms with high probability). This is consistent with the message generation interval calculated using the SAE congestion control in Section~\ref{sec_sub_cc} for a vehicle density of 400 VUE/km.

From these figures, it can be seen that the one-shot transmissions do indeed improve the tail behavior of the IPG CCDF in all cases (i.e., they lead to a steeper drop in the tail). We measure this gain by calculating the average relative improvement in the IPG CCDF for all IPG values in the interval $[3,\,10]$ seconds. For example, let $F_{\mathrm{s}}(i)$ be the CCDF of the IPG evaluated at $i$ milliseconds of the simulation scenario $s$. The relative improvement in the IPG tail of simulation scenario \textit{26-20} is calculated as follows:
\begin{equation}
    \delta_{i}=\frac{F_{\mathrm{OFF-20}}(i)-F_{\mathrm{26-20}}(i)}{F_{\mathrm{OFF-20}}(i)}.
\end{equation}
The average relative improvement is then calculated as the average of $\{\delta_i\}_{i \in \mathcal{I}}$, where $\mathcal{I}$ is a quantization of the interval $[3,10]$ seconds with steps of size 1 ms. As shown in Fig.~\ref{fig_ipgccdf}(a), at 20 MHz bandwidth, the one-shot transmissions with $C_{\mathrm{o}}$ configurations of $[5,\,15]$, and $[2,\,6]$ give an average relative improvement at distance bin 200 m of $\approx$ 94.7$\%$ and 99.6$\%$, respectively. The tail behavior improvement is slightly lower for 10 MHz bandwidth. Essentially, increasing the number of one-shot reselections ($[2,\,6]$ vs. $[5,\,15]$) decreases the number of instances with long IPGs (i.e., persistent BSM collisions) and improves the IPG tail behavior. 

\begin{table}
\renewcommand{\arraystretch}{0.75}
  \centering
  \caption{Relative gains in inter-packet gap statistics.}\label{tab_ipgstats}
        \begin{tabular}{ P{2cm} P{2cm} P{2cm} P{2cm} P{2cm}}
        \hline
        \textbf{V2V distance}  & \textbf{515-20} & \textbf{26-20} & \textbf{515-10} & \textbf{26-10}\\ \hline
        \multicolumn{5}{c}{\textbf{Average improvement of IPG CCDF}}\\ \hline
        200 & .94731 & .99638 & .93058 & .99298\\ \hline
        300 & .90240 & .98827 & .84946 & .97370\\ \hline
        400 & .82197 & .96504 & .73386 & .92858\\ \hline
        \multicolumn{5}{c}{\textbf{Improvement of IPG 99.9th-percentile}}\\ \hline
        200 & .09445 & .31926 & .37352 & .54761\\ \hline
        300 & .36184 & .49326 & .39716 & .54189\\ \hline
        400 & .34182 & .51493 & .35230 & .53000\\ \hline
        \end{tabular}
        \vspace{-.25in}
\end{table}

Fig.~\ref{fig_ipgccdf} also demonstrates that the 20 MHz configurations have better tail behavior than their counterparts with 10 MHz. This is because the higher number of available resources leads to a lower collision probability. It is worth noting that, at a V2V distance bin of 200 m, the IPG tail performance of the 10 MHz with $C_{\mathrm{o}}\in[2,\,6]$ is better than that of the 20 MHz case without one-shot transmissions and with $C_{\mathrm{o}}\in[5,\,15]$. In other words, the negative impacts of using smaller bandwidths can be offset by using rapid one-shot reselections (i.e., the $[2,\,6]$ configuration). Fig.~\ref{fig_ipgccdf}(c) also reveals a smaller improvement in the IPG tails when Tx-Rx pairs are further separated from each other (i.e., bin 400 vs. 200 m). As the V2V separation between Tx-Rx pairs increases, the number of potential interferers increases leading to longer IPGs due to the higher number of persistent collisions. Hence, one-shot transmissions are less likely to stop long IPGs with a high number of interferers.

Fig.~\ref{fig_ipg999} provides another way to see the advantage of using the one-shot transmissions to improve the IPG tail in C-V2X networks. Specifically, we compare the length of IPGs that are larger than 99.9\% of the recorded IPGs at a given V2V distance. Again, each distance bin represents 50 m. As shown in Fig.~\ref{fig_ipg999}, using $[5,\,15]$ and $[2,\,6]$ configurations of $C_{\mathrm{o}}$ with 20 MHz bandwidth improves the 99.9th-percentile of IPG by $\approx$ 9.4\%, and 31.9\% at distance bin 200 m, and by $\approx$ 36.2\% and 49.3\% at distance bin 300 m, respectively. A significant improvement is also observed at 10 MHz when one-shot transmissions are used. The relative improvements in the IPG statistics are summarized in Table~\ref{tab_ipgstats}.

\subsection{Information Age (IA)}\label{sec_sub_ia}

\begin{figure}
  \begin{center}
  \includegraphics[width=8cm,height=8cm,,keepaspectratio]{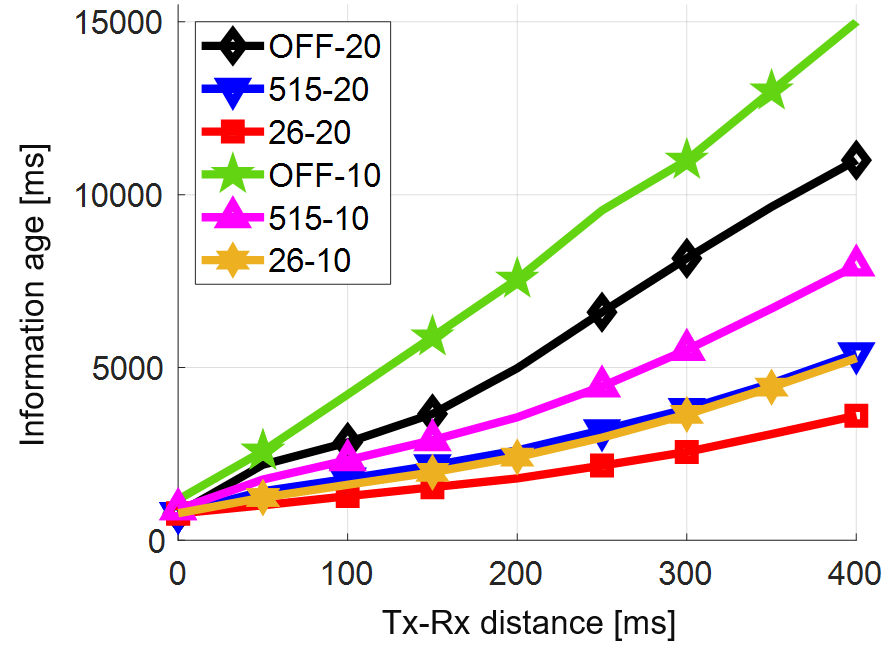}
  \caption{99.9th-percentile of information age.}\label{fig_ia999}
  \vspace{-.25in}
  \end{center}
\end{figure}
Another important metric in V2X networks which is related to IPG is IA. This metric measures the time elapsed since the generation of the latest successfully received BSM at destination VUEs~\cite{5gaa_PerforTest}. The IA of a BSM received at VUE B from VUE A at $t_{\mathrm{c}}$ is defined as 
    $I_{\mathrm{B,A}}=t_{\mathrm{c}}-t_{\mathrm{s,B}}+\eta_{\mathrm{A}}$,
where $t_{\mathrm{c}}$ and $t_{\mathrm{s, B}}$ denote the current time and the time of the last successfully received BSM at VUE B, respectively. $\eta_{\mathrm{A}}$ denotes the application-physical layer latency of VUE A, and is defined as $\eta_{\mathrm{A}}=t_{\mathrm{g,A}}-t_{\mathrm{r,A}}$, where $t_{\mathrm{g,A}}$ and $t_{\mathrm{r,A}}$ denote the BSM generation time and the actual transmission time at VUE A, respectively.  Here, we use a \textit{uniform} time sampling of the IA sawtooth sample path to collect the IA data based on time samples of the received BSMs (i.e., peak ages). In particular, the IA CCDF $F(i)$ can be interpreted as the fraction of time that the IA exceeds $i$ milliseconds. Fig.~\ref{fig_iaccdf}(a) shows that the $[5,\,15]$ and $[2,\,6]$ one-shot configurations with 20 MHz improve the IA tail by $\approx$ 98.1$\%$ and 99.9$\%$ at V2V distance bin 200 m, respectively, where the improvement is calculated using the same approach as we used for the IPG tail.

Fig.~\ref{fig_iaccdf} also reveals that these results are consistent with Figs.~\ref{fig_ipgccdf} and~\ref{fig_ipg999}. Specifically, IA performance of the $[2,\,6]$ configuration at 10 MHz outperforms that of the 20 MHz without one-shot transmissions and with the $[5,\,15]$ configuration. Further, it demonstrates that using one-shot transmissions yields the best improvement in IA performance at a distance bin 200 m (vs. all other bins). Fig.~\ref{fig_ia999} confirms the IA tail improvement when one-shot reselection is used. Specifically, the $[2,\,6]$ configuration with 20 MHz improves the 99.9th-percentile of IA at distance bins 200 and 300 m by $\approx$ 64\% and 68.6\%, respectively. Table~\ref{tab_iastats} summarizes the IA CCDF tail and the 99.9th-percentile relative gains. 
\begin{table}
\renewcommand{\arraystretch}{0.75}
  \centering
  \caption{Relative gains in IA statistics.}\label{tab_iastats}
        \begin{tabular}{ P{2cm} P{2cm} P{2cm} P{2cm} P{2cm}}
        \hline
        \textbf{V2V distance}  & \textbf{515-20} & \textbf{26-20} & \textbf{515-10} & \textbf{26-10}\\ \hline
        \multicolumn{5}{c}{\textbf{Average improvement of IA CCDF}}\\ \hline
        200 & .98139 & .99925 & .96970 & .99780\\ \hline
        300 & .95974 & .99670 & .92808 & .99139\\ \hline
        400 & .91295 & .98884 & .85946 & .97371\\ \hline
        \multicolumn{5}{c}{\textbf{Improvement of IA 99.9th-percentile}}\\ \hline
        200 & .47741 & .64003 & .52891 & .68117\\ \hline
        300 & .53449 & .68643 & .49891 & .66882\\ \hline
        400 & .50273 & .67209 & .46987 & .64860\\ \hline
        \end{tabular}
        \vspace{-.25in}
\end{table}

\subsection{Packet Reception Ratio (PRR)}\label{sec_sub_prr}
\begin{figure}
  \begin{center}
  \includegraphics[width=8cm,height=8cm,,keepaspectratio]{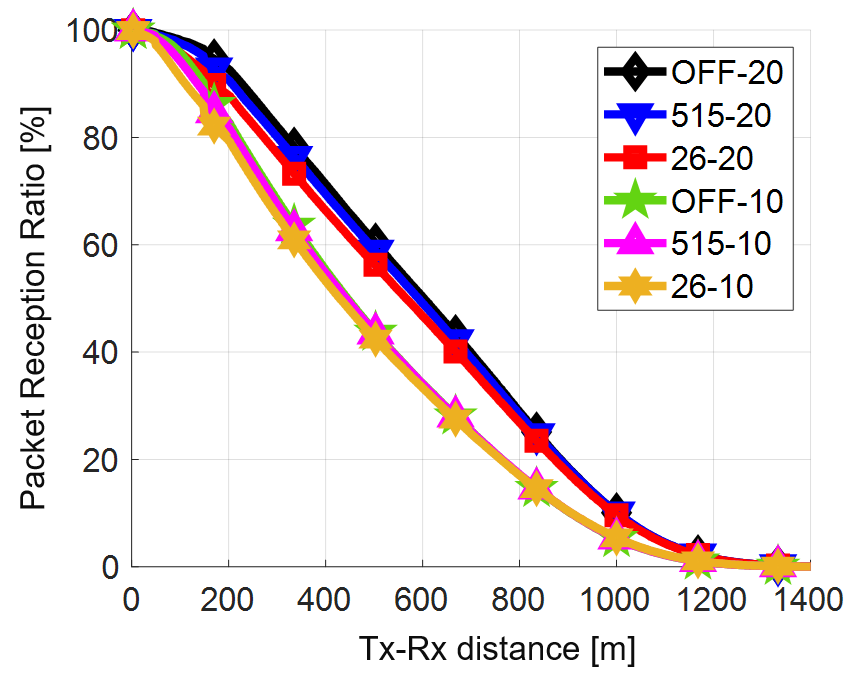}
  \caption{Packet reception ratio.}\label{fig_prr}
  \vspace{-.25in}
  \end{center}
\end{figure}
Next, we analyze the PRR performance when one-shot reselection is used. PRR is used to evaluate the transmission reliability in C-V2X networks~\cite{3gpp37885}. In particular, PRR at a distance bin $d$ is calculated by $R/T$, where $T$ denotes the total number of transmitted BSMs between all Tx-Rx pairs that are separated by $d$, and $R$ is the number of successfully received BSMs among $T$. Fig.~\ref{fig_prr} shows the PRR with respect to the V2V distance where each distance bin represents 1 m. As expected, the one-shot configurations at 20 MHz outperform those at 10 MHz because of the higher number of available resources. Fig.~\ref{fig_prr} also reveals that the improvement in IPG and IA tails comes at the expense of slight degradation in PRR performance. Specifically, PRR performance degrades by $\approx$ 2.2\% and 5.3\% at distance bin 200 m when one-shot $[5,\,15]$ and $[2,\,6]$ configurations are used, respectively, with 20 MHz.

Similarly, PRR performance at the same distance bin drops by $\approx$ 1.4\% and 4.2\% when the same configurations are used with 10 MHz. Table~\ref{tab_prrstats} summarizes the relative change in PRR statistics for different simulation scenarios. It is somewhat expected that since one-shot transmissions introduce more randomness into the SPS transmissions they will lead to a lower PRR, which this result confirms. However, as shown, this degradation is relatively small and may be acceptable in order to get the improvement in IPG and IA tails.

\begin{table}
\renewcommand{\arraystretch}{0.75}
  \centering
  \caption{Relative change in PRR statistics.}\label{tab_prrstats}
        \begin{tabular}{ P{2cm} P{2cm} P{2cm} P{2cm} P{2cm}}
        \hline
        \textbf{V2V distance}  & \textbf{515-20} & \textbf{26-20} & \textbf{515-10} & \textbf{26-10}\\ \hline
        200 & -.02202 & -.05279 & -.01369 & -.04207\\ \hline
        300 & -.02373 & -.06055 & -.01401 & -.04296\\ \hline
        400 & -.02545 & -.06851 & -.01127 & -.03814\\ \hline
        \end{tabular}
        \vspace{-.25in}
\end{table}

\subsection{PSSCH Channel Busy Ratio (CBR)}\label{sec_sub_cbr}
\begin{figure}
  \begin{center}
  \includegraphics[width=8cm,height=8cm,,keepaspectratio]{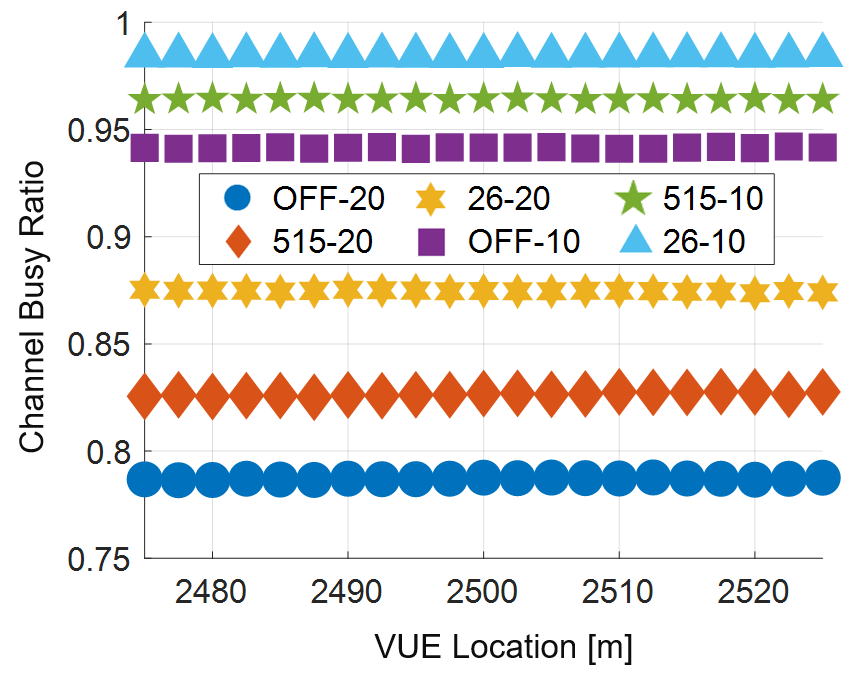}
  \caption{Channel busy ratio.}\label{fig_cbr}
  \vspace{-.25in}
  \end{center}
\end{figure}

Introducing an additional degree of randomness into the SPS process is also expected to decrease the time interval of reserving the same set of VRBs for consecutive BSM transmissions. Such a degradation may lead to a surge in the occupied number of wireless resources creating a highly congested scenario. Hence, it is essential to evaluate the utilization of PSSCH VRBs when one-shot transmissions are used~\cite{3gpp36214}. We do so by using the CBR, which is defined as $X/Y$, where $Y$ denotes the total number of VRBs in the SPS selection window (i.e., the most recent 100 ms), and $X$ denotes the VRBs \textit{among $Y$} whose average sidelink RSSI (S-RSSI) exceeds a threshold of -94 dBm at a given VUE~\cite{etsi}. The average S-RSSI per VRB and the instantaneous CBR per VUE are updated periodically once per sub-frame (i.e., 1 ms) and measured over a sliding window of 1000 ms and 100 ms, respectively. Fig.~\ref{fig_cbr} shows the average CBR of VUEs located within 25 m from the center of the highway. As shown, CBR dropped from $\approx$ 94.2\% to 78.8\% when 20 MHz is used (vs. 10 MHz) without one-shot transmissions because of the higher number of available resources for the same number of VUEs.  

Fig.~\ref{fig_cbr} also reveals that the $[5,\,15]$ one-shot configuration increases CBR from 94.2\% to 96.5\%, and from 78.8\% to 82.7\% at 10 and 20 MHz bandwidths, respectively. Similarly, the $[2,\,6]$ configuration increases CBR to 98.4\% and 87.5\% at 10 and 20 MHz, respectively. This is because the average S-RSSI per VRB is measured over a sliding window of the past 1000 sub-frames not only the current sub-frame. Essentially, using the one-shot transmissions decreases the resource reservation interval, during which a VUE keeps using the same VRBs for BSM transmission. Consequently, the total number of occupied VRBs (i.e., VRBs whose average S-RSSI exceeds -94 dBm) in the most recent 1000 ms increases as the number of one-shot reselections increases (see $[2,\,6]$ vs. $[5,\,15]$ configurations in Fig.~\ref{fig_cbr}). It is worth mentioning that the CBR statistics (with and without one-shot) could have been essentially the same if the S-RSSI per VRB was measured using only the most recent sub-frame (i.e., 1 ms). This is because the SPS-granted VRBs, of the current sub-frame, are considered free resources if a given VUE decides not to use them and does one-shot transmission using other VRBs.   
\subsection{Performance Evaluation of Other Densities}\label{sec_sub_cbr}
\begin{figure}
  \begin{center}
  \includegraphics[width=12cm,height=12cm,,keepaspectratio]{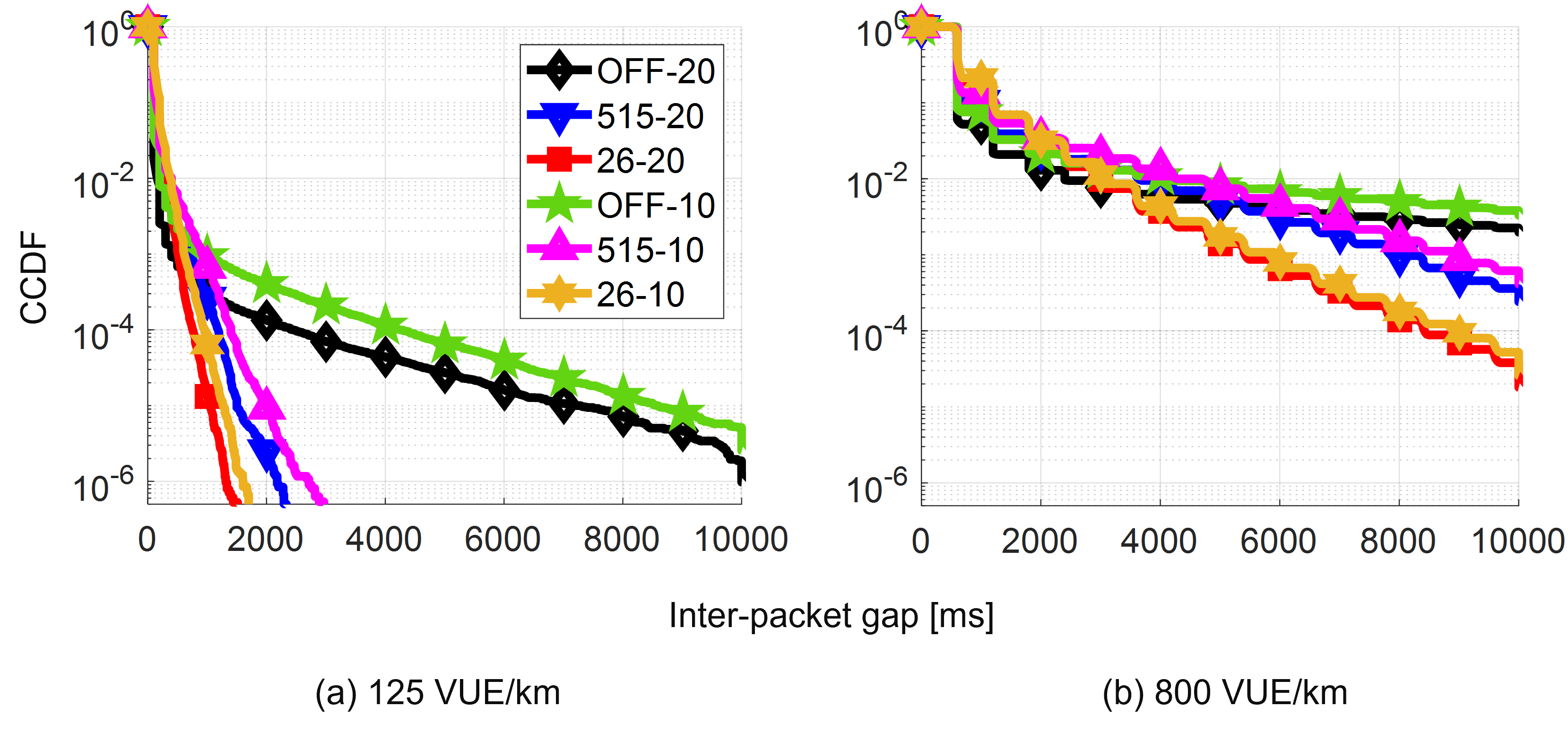}
  \caption{IPG of distnace bin 200 m at different densities.}\label{fig_ipg6254000}
  \vspace{-.25in}
  \end{center}
\end{figure}

\begin{figure}
  \begin{center}
  \includegraphics[width=12cm,height=12cm,,keepaspectratio]{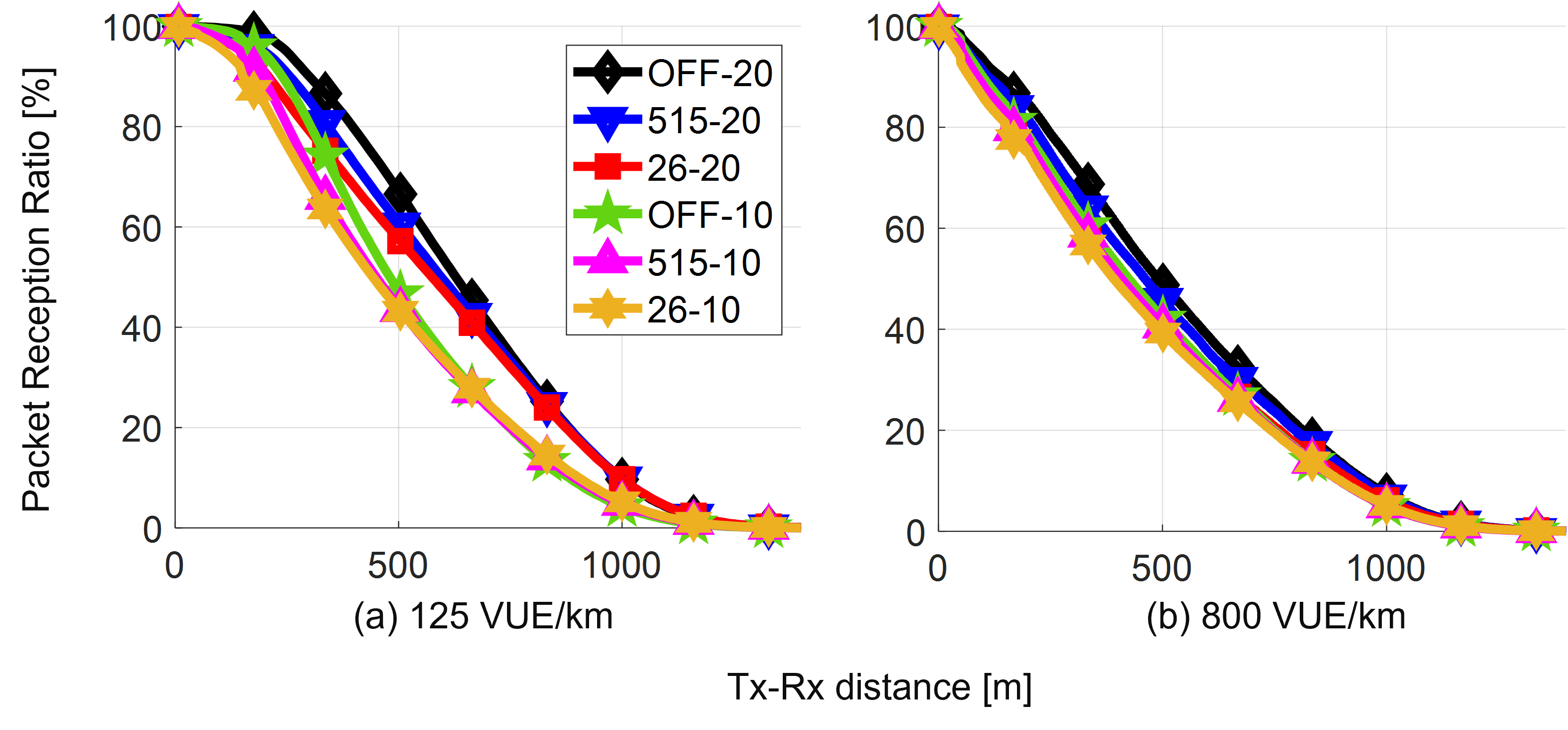}
  \caption{PRR statistics at different densities.}\label{fig_prr625400}
  \vspace{-.25in}
  \end{center}
\end{figure}

In this section, we evaluate the performance of the IPG tail and PRR with other vehicle densities. Specifically, we use the 125 and 800 VUE/km densities which correspond to BSM generation intervals of 100 and 600 ms, respectively. Here, we calculate the performance change using the same approaches as in Sections~\ref{sec_sub_ipg} and~\ref{sec_sub_prr}. Fig~\ref{fig_ipg6254000}(a) shows that the IPG tails of all one-shot configurations are improved by almost 100\% at 125 VUE/km. Alternatively, Fig.~\ref{fig_ipg6254000}(b) shows smaller gains at 800 VUE/km. Essentially, increasing the number of VUEs decreases the number of available VRBs increasing the collision probability. Hence, using one-shots at high densities may not lead to a significant IPG tail improvement because it is more likely that a given VUE will do a one-shot transmission resulting in another collision.

Fig.~\ref{fig_ipg6254000} also suggests that using the $[2,\,6]$ configuration of one-shot transmissions (at both bandwidths) with 800 VUE/km may slightly decrease the IPG tail performance for all IPGs $<\xi$, where $\xi=3$ seconds, because of the higher number of collisions. This threshold decreases as the vehicle density decreases because of the lower number of collisions. In particular, $\xi=1$ and $0.5$ second at 400 and 125 VUE/km, respectively. The same trend is observed for the $[5,\,15]$ one-shot configuration with slightly different thresholds. We point out that all the above observations also apply to the other V2V distance bins and IA statistics. Fig.~\ref{fig_prr625400} shows that using the $[2,6]$ configuration of one-shot transmissions at 20 MHz bandwidth and 200 m V2V distance bin results in the highest PRR drop of $\approx$ 9\% at 800 VUE/km. The PRR drop is smaller at lower densities (i.e., PRR dropped by $\approx$ 7\% using the same one-shot settings at 125 VUE/km). As expected, the highest PRR drop happens at the highest vehicle density when one-shot transmissions are used because of the smaller number of available VRBs. Table~\ref{tab_prr6254000stats} summarizes the relative change in the PRR statistics for different vehicle densities.

\begin{table}
\renewcommand{\arraystretch}{0.75}
  \centering
  \caption{PRR relative change at different densities.}\label{tab_prr6254000stats}
        \begin{tabular}{ P{2cm} P{2cm} P{2cm} P{2cm} P{2cm}}
        \hline
        \textbf{V2V distance}  & \textbf{515-20} & \textbf{26-20} & \textbf{515-10} & \textbf{26-10}\\ \hline
        \multicolumn{5}{c}{\textbf{125 VUE/km}}\\ \hline
        200 & -.03384 & -.07442 & -.05684 & -.10525\\ \hline
        300 & -.05141 & -.11899 & -.11064 & -.14846\\ \hline
        400 & -.08583 & -.14409 & -.09817 & -.11999\\ \hline
        \multicolumn{5}{c}{\textbf{800 VUE/km}}\\ \hline
        200 & -.03827 & -.09392 & -.01981 & -.04361\\ \hline
        300 & -.06395 & -.14728 & -.03007 & -.05943\\ \hline
        400 & -.06647 & -.17274 & -.03544 & -.06939\\ \hline
        \end{tabular}
        \vspace{-.25in}
\end{table}

\subsection{Robustness of the One-shot Transmissions Method}\label{sec_sub_alt}
\begin{figure}
  \begin{center}
  \includegraphics[width=8cm,height=8cm,,keepaspectratio]{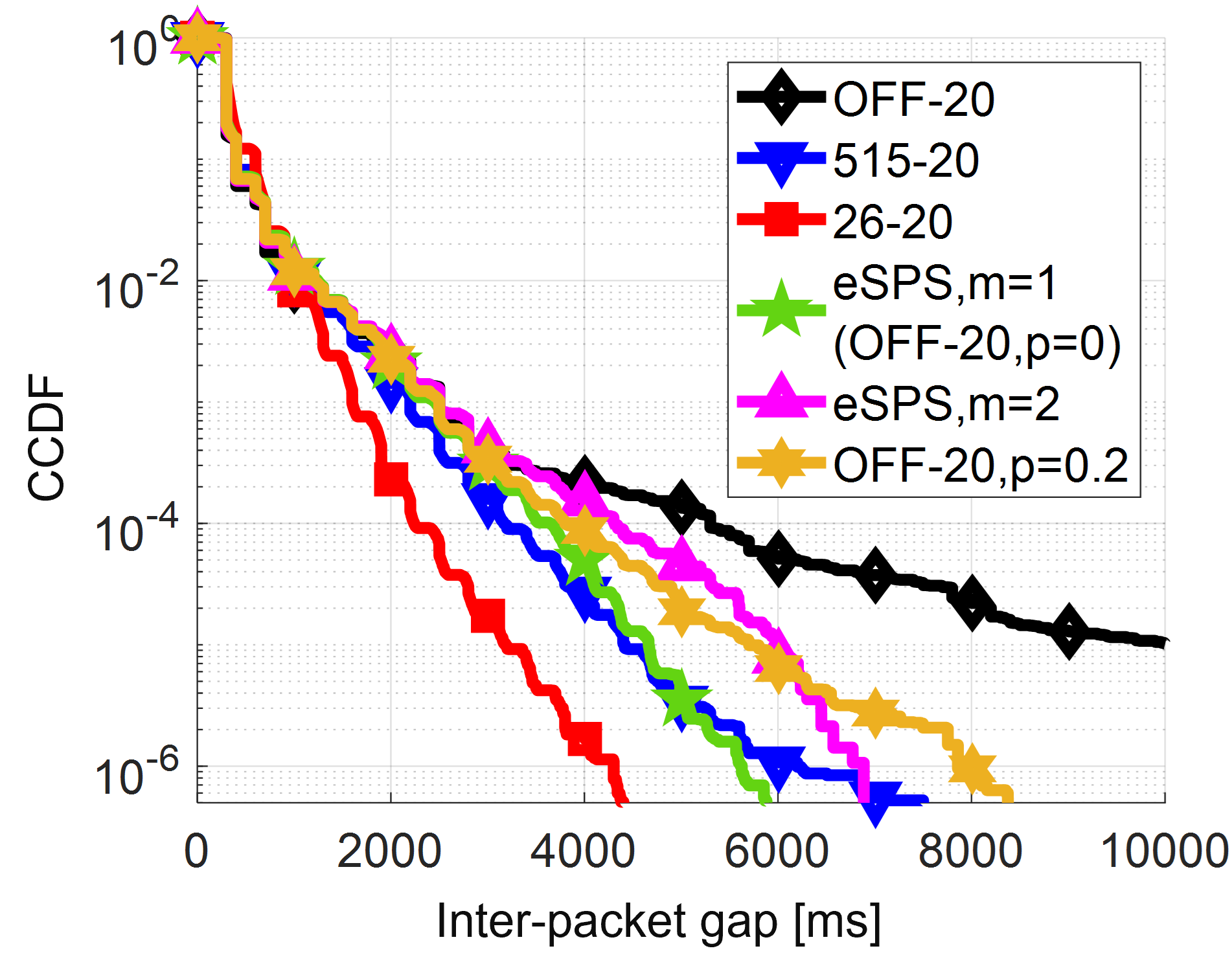}
  \caption{Alternative methods for improving IPG tail.}\label{fig_altmthdsIPG}
  \vspace{-.25in}
  \end{center}
\end{figure}

In this section, we investigate the advantage of using the one-shot procedure against some alternative techniques to improve the IPG/IA distribution tails in C-V2X networks. Specifically, the one-shot performance is compared with (1) the modified SPS scheme that has been presented in~\cite{bspots}, and (2) the legacy SPS with lower values for the probability of keeping the current VRBs for the next SPS intervals (i.e., ${p}_{\mathrm{p}}$). The authors in~\cite{bspots} presented an enhanced SPS to reduce the duration of using the same interfered VRBs for multiple SPS intervals. This method aims to decrease the probability of losing consecutive BSMs without impacting transmission reliability. In particular, a BSM can be sent using the same set of VRBs for a maximum of arbitrary $m$ SPS intervals. Note that an SPS interval consists of $C_{\text{s}}$ BSM transmissions which is chosen uniformly at random between $\left[\alpha,\,\beta\right]$. Once the SPS reselection counter (i.e., $C_{\mathrm{s}}$) expires at the $m^{\text{th}}$ SPS interval, new VRBs are reselected regardless of the value of ${p}_{\mathrm{p}}$.

\begin{figure}
  \begin{center}
  \includegraphics[width=8cm,height=8cm,,keepaspectratio]{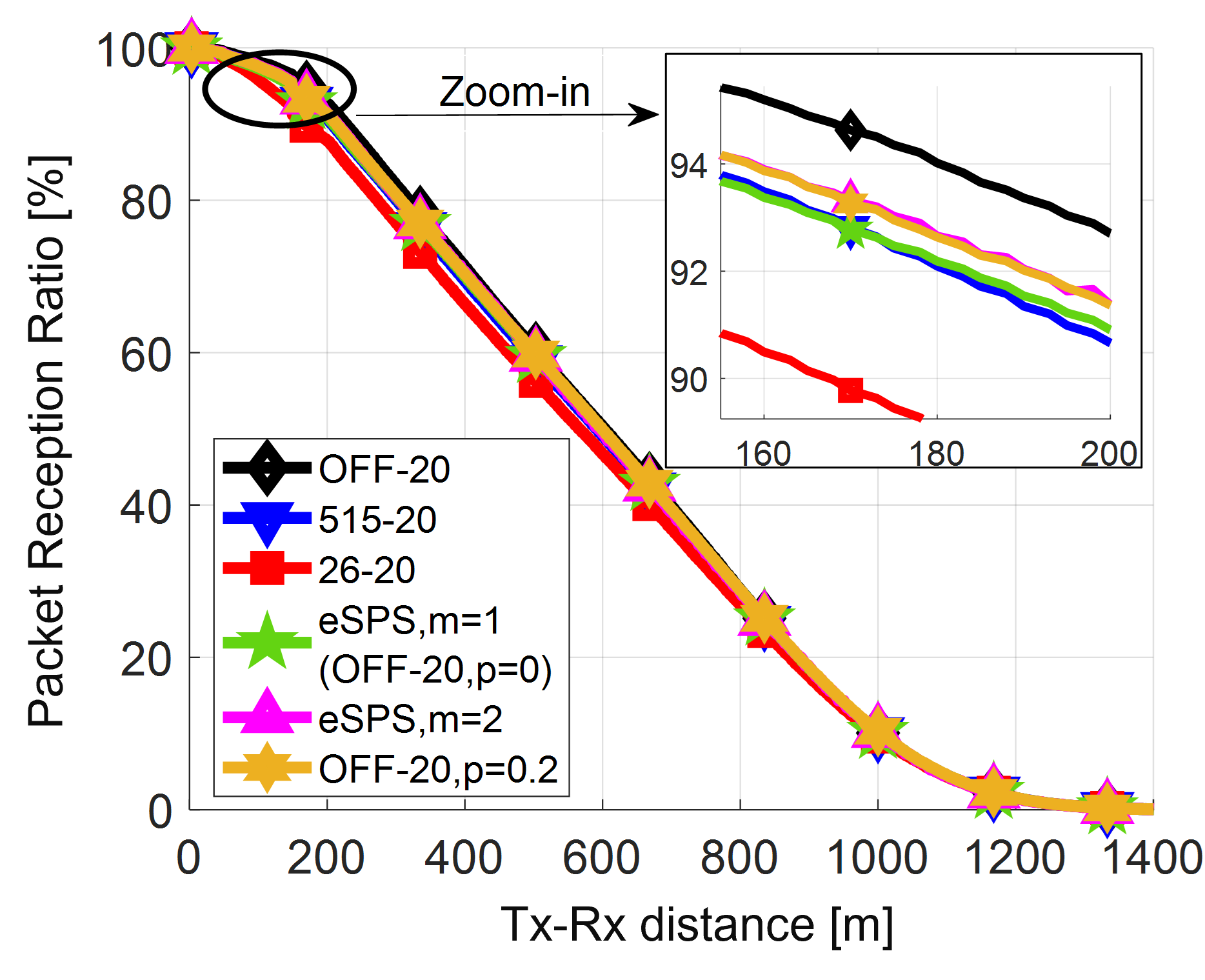}
  \caption{Alternative methods: PRR performance.}\label{fig_altmthdsPRR}
  \vspace{-.25in}
  \end{center}
\end{figure}

Fig.~\ref{fig_altmthdsIPG} shows the related IPG CCDF tail improvement of different configurations of the modified SPS method in~\cite{bspots} which we refer to as \textit{eSPS}. As shown, the IPG tail improvement increases with lower values of $m$ because the time a VUE stays in a given choice of colliding resources decreases as $m$ decreases. Fig.~\ref{fig_altmthdsIPG} reveals that the best achievable IPG tail improvement using eSPS (when $m=1$) is similar to that of the $[5,\,15]$ configuration of interleaved SPS and one-shot transmissions. It also demonstrates that the $[2,\,6]$ configuration significantly outperforms eSPS with $m=1$ which confirms the superiority of the interleaved one-shot SPS technique among the state-of-the-art SPS-based methods to improve the IPG/IA tail behaviors in C-V2X networks. Another potential method to decrease the duration of consecutive BSM losses (due to the use of the same interfered VRBs for multiple SPS intervals) is to use lower values for ${p}_{\mathrm{p}}$. Essentially, decreasing ${p}_{\mathrm{p}}$ increases the probability of reselections at the end of the SPS intervals (i.e., when $C_{\mathrm{s}}$ expires). As shown in Fig.~\ref{fig_altmthdsIPG}, the best IPG tail improvement using such a method can be achieved when ${p}_{\mathrm{p}}=0$. It is worth noting that using ${p}_{\mathrm{p}}=0$ with the legacy SPS procedure is equivalent to using $m=1$ with the eSPS method. In that, the resource reselection is triggered once $C_{\mathrm{s}}$ expires.  

Fig.~\ref{fig_altmthdsPRR} confirms that using both the legacy SPS method with different values of ${p}_{\mathrm{p}}$ and the eSPS technique to improve the IPG tail behavior does not impact the BSM transmission reliability. As shown, introducing more randomness into the legacy SPS procedure (by either lowering the ${p}_{\mathrm{p}}$ or using the eSPS method) slightly degrades the PRR performance. A such trade-off has been observed when one-shot is used giving these methods no edge against the interleaved SPS one-shot technique. As discussed earlier, this degradation is small compared with the achievable improvement in the IPG and IA tails. It is worth mentioning that the results in this section are generated for the medium vehicle density (400 VUE/km) at the 20 MHz bandwidth configuration. Similar trends are observed at the 10 MHz bandwidth configuration, different vehicle densities, and different values of $m$ and ${p}_{\mathrm{p}}$ parameters. 

\section{Inter-packet Gap Analysis}\label{sec_analmodel}

In this section, we propose an analytical model to estimate IPG tail behavior. In particular, we use a non-uniform geometric distribution-based model to approximate the slope of the IPG CCDF tails on a log scale, under different resource reselection modes (i.e., SPS with and without one-shot transmissions), vehicle densities, and bandwidth configurations. The proposed model is validated by comparing its results with that obtained using the system-level Monte Carlo simulations. For a given transmitter-receiver pair, conditioned on a first BSM transmission not being received, let $T$ be a random variable that indicates the IPG length in terms of the required number of BSM transmission opportunities until a successful transmission. In other words, if $T>k$ for some non-negative integer $k$, the transmitter has sent $k$ consecutive BSMs that have not been correctly received at the receiving VUE.  

To develop our analytical model, we assume that if a BSM is not received, it is due to interference from one dominant interferer that is using an overlapping VRB with the transmitter\footnote{This assumption is based on the data extracted from extensive simulations for different bandwidth configurations, vehicle densities, and radio resource reselection strategies. The data show that packet collisions between the transmitter and the interferer are the main reason for long IPGs. Other reasons for long IPGs such as HD transmissions, deep fading, and low levels of signal power below the thermal noise happen in very few instances (i.e., rare samples) which can be fairly ignored.}. In this case, under SPS scheduling, this interference will persist until either the transmitter or the receiver reselects another resource or does a one-shot transmission. To begin, we give a non-uniform geometric model that can apply to either the SPS or one-shot reselection process. For either of these, we assume that the probability of a resource reselection during the $k^{\mathrm{th}}$ BSM transmission opportunity is given by 
$q^{(k)}(\rho,\sigma,p),$ independently over time. Here, $\rho$ and $\sigma$ denote the limits of the reselection window and $p$ represents the reselection probability when the reselection counter expires. Hence, the SPS-based and one-shot-based reselection probabilities can be denoted as $q_\mathrm{s}^{(k)}=q^{(k)}(\alpha_\mathrm{s},\,\beta_\mathrm{s},\,p_{\mathrm{r}})$ and 
$q_\mathrm{o}^{(k)}=q^{(k)}(\alpha_\mathrm{o},\,\beta_\mathrm{o},\,1)$, respectively. Given that reselection is only triggered when the reselection counter expires, there can be no reselections in the first $\rho$ BSM transmission opportunities. Hence, we take this into account and set $q^{(k)} = 0$ for $k \leq \rho$. Further, for $k > \rho$, we set $q^{(k)}$ so that for large $k$, the expected number of reselections after $k$ opportunities are equal to $k/E(T_r)$ where $E(T_r)$ denotes the expected time between reselections and is given by: 
\begin{equation}
E(T_r) = \left(\rho + \frac{\sigma - \rho}{2}\right) \frac{1}{p_s}.
\end{equation}
In that, we want to select $q^{(k)}$ to satisfy:
\begin{equation}
q^{(k)} (k-\rho) = \frac{2kp_s}{\sigma+ \rho}.
\end{equation}
Hence, 
\begin{equation}
q^{(k)} = \frac{2kp_s}{(\sigma+ \rho)(k-\rho)}.
\end{equation}
Now, let $q_{\mathrm{\mathrm{i}}}^{(k)}$ represent the reselection probability of the interleaved one-shot SPS processes. Assuming that the two processes are independent, $q_{\mathrm{\mathrm{i}}}^{(k)}$ is given by: 
\begin{equation}\label{eq_qe}
    q_{\mathrm{\mathrm{i}}}^{(k)}=q_{\mathrm{o}}^{(k)}(1-q_{\mathrm{s}}^{(k)})+q_{\mathrm{s}}^{(k)}(1-q_{\mathrm{o}}^{(k)})+q_{\mathrm{s}}^{(k)}q_{\mathrm{o}}^{(k)}.
\end{equation}
In that, the first and second terms represent the event when the reselection is triggered by the one-shot and the SPS processes, respectively. The last term represents the event when the SPS and one-shot reselection counters expire simultaneously. Note that $ q_{\mathrm{\mathrm{i}}}^{(k)}$ can be used for scenarios without one-shot transmission to denote the reselection probability of only the SPS process as $q_{\mathrm{\mathrm{i}}}^{(k)}=q_{\mathrm{\mathrm{s}}}^{(k)}$. The probability of a successful BSM transmission (due to a reselection) after $k$ transmission opportunities (i.e., failures) can then be calculated as follows:
\begin{equation}\label{eq_ps_tr}
    p_\mathrm{u}^{(k)}=(1-q_{\mathrm{i}}^{(k)})q_{\mathrm{i}}^{(k)}P_{\mathrm{f}}+(1-q_{\mathrm{i}}^{(k)})q_{\mathrm{i}}^{(k)}+{\left(q_{\mathrm{i}}^{(k)}\right)}^2P_{\mathrm{f}},
\end{equation}
where the first term represents the probability that the interferer continues using the same VRBs with probability $\left(1-q_{\mathrm{i}}^{(k)}\right)$, and the transmitter reselects with probability $q_{\mathrm{i}}^{(k)}$ and that reselection results in a successful transmission with probability $P_{\mathrm{f}}$ (i.e., it does not reselect into the same VRBs as another strong interferer). We assume that the probability of successful transmission because of one-shot reselection is the same as the average success probability of any transmission\footnote{This assumption is confirmed by collecting separate PRR statistics for SPS-based and one-shot-based BSM transmissions using extensive system-level simulations. Results show that the two methods have an approximately identical PRR behavior.}. The second term represents the opposite scenario, in which, the transmitter stays using the same VRBs with probability $\left(1-q_{\mathrm{i}}^{(k)}\right)$ and the interferer reselects with probability $q_{\mathrm{i}}^{(k)}$. In this case, there is no $P_{\mathrm{f}}$ term as we are assuming a single dominant interferer and a new dominant interferer would not select occupied VRBs. Finally, the third term represents that both the transmitter and receiver reselect new VRBs. 

\begin{figure}
  \begin{center}
  \includegraphics[width=8cm,height=8cm,,keepaspectratio]{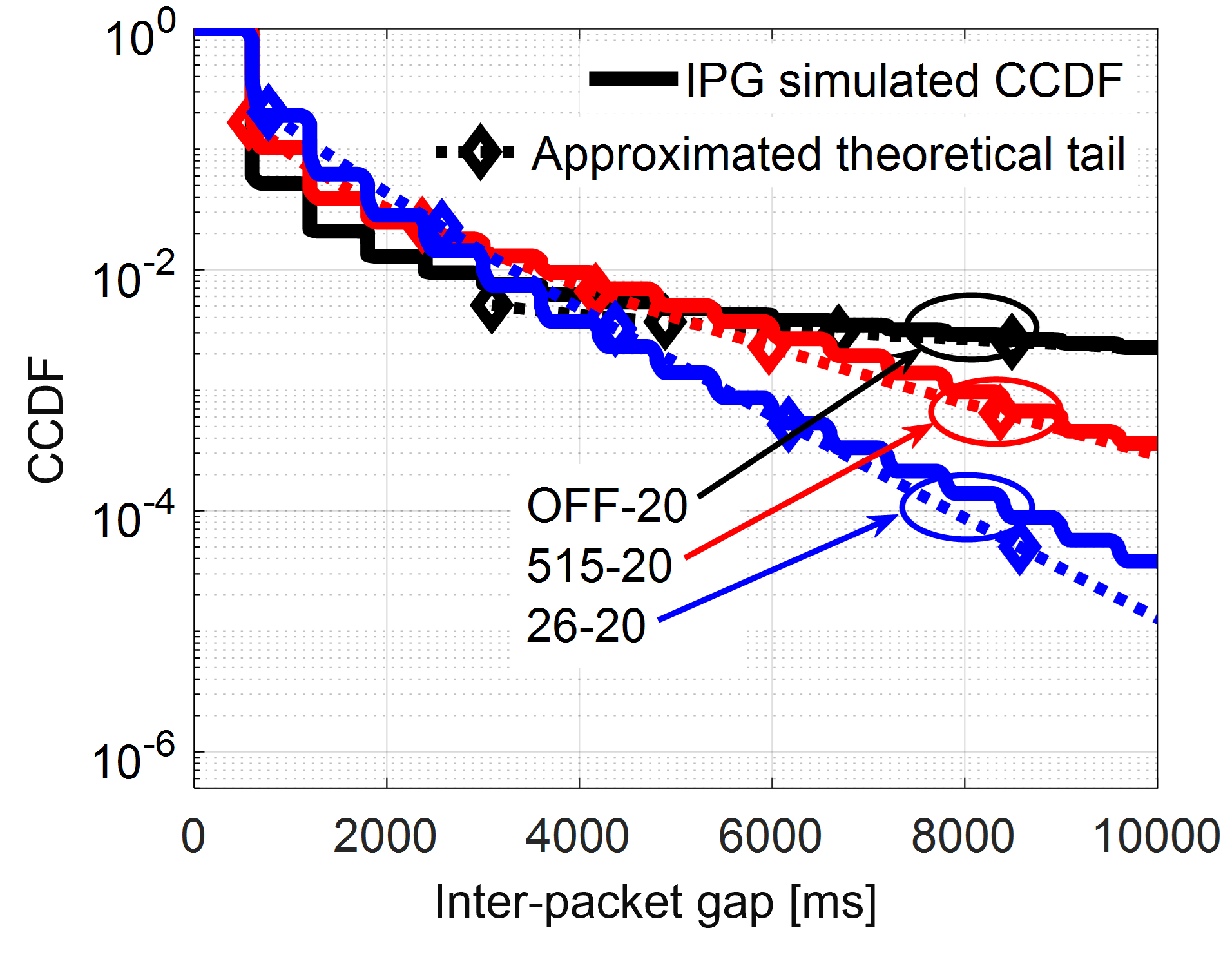}
  \caption{Simulated versus approximated IPG tails at 200 m.}\label{fig_valid_norc}
  \vspace{-.25in}
  \end{center}
\end{figure}

Hence, the probability that the IPG exceeds $k$ BSM transmission opportunities can be calculated as:
\begin{equation}\label{eq_tk}
    P(T>k)=\prod^{k}_{i=1}(1-p_{\mathrm{u}}^{(i)}).
\end{equation}
We then use~\ref{eq_tk} to approximate the slope of IPG CCDF tail\footnote{Note that since $T$ is defined conditioned on one BSM not arriving, then there will be a constant offset between $P(T>t)$ and the probability that the IPG exceeds the same amount, which corresponds to the probability that the first BSM does not arrive. Hence, this captures the slope of the IPG CCDF but not the actual value.} and compare it with that obtained via the system-level simulations as presented in Fig.~\ref{fig_valid_norc}. Here, the dotted lines represent the slope for the analytical results, and the solid lines represent the simulation results. The IPG slope approximations are for three configurations of $C_{\mathrm{o}}$, OFF, $[2,\,6]$, and $[5,\,15]$ at 20 MHz with 800 VUE/km. Here, $P_{\mathrm{f}}$ is calculated using Fig.~\ref{fig_prr} based on the operating scenario and the desired V2V distance bin. We run the simulations for 500 seconds ($\approx\mathrm{10^8}$ samples) to obtain the numerical results with a good confidence level. It can be seen that the IPG tail approximation in~(\ref{eq_tk}) is reasonable for the different reselection modes. Note that the minor deviations between the analytical and simulated slopes at small IPG values (i.e., $\le$ 3 sec) are because these approximated slopes are essentially derived for large IPG values. Further, this model does not consider other factors that may result in small IPGs such as IBE and HD transmission. 
Similar conclusions hold for other distance bins, bandwidths, and vehicle densities as long as the BSM generation interval is a multiple of 100 ms. 

\begin{figure}
  \begin{center}
  \includegraphics[width=8cm,height=8cm,,keepaspectratio]{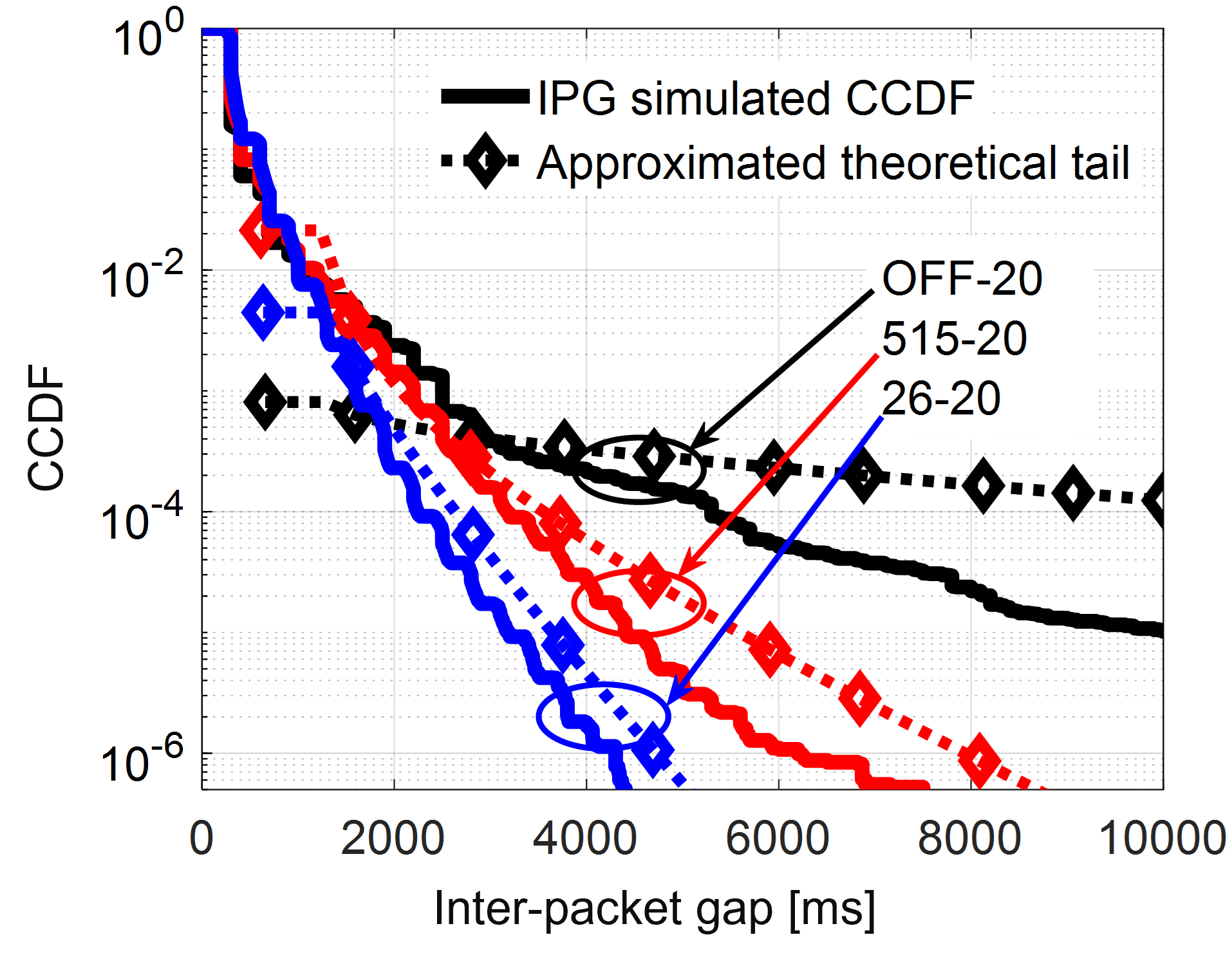}
  \caption{Approximated IPG tails without considering slippage.}\label{fig_noslip}
  \vspace{-.25in}
  \end{center}
\end{figure}

Fig.~\ref{fig_noslip} shows a similar comparison using 400 VUE/km at 20 MHz.  In this case, due to congestion control, the BSM generation interval is not an integer multiple of 100 ms. As shown, there is a noticeable gap between the approximations in (\ref{eq_tk}) and the simulated results.  This is because the reselection of new VRBs is no longer the only dominant reason for ending IPGs. In particular, it is now likely that an IPG ends because the transmitter's and the interferer's transmissions get out of phase. We refer to this scenario as {\it slippage} and explain it next using the following numerical example. 

\begin{figure}
  \begin{center}
  \includegraphics[width=12cm,height=8.85cm,,keepaspectratio]{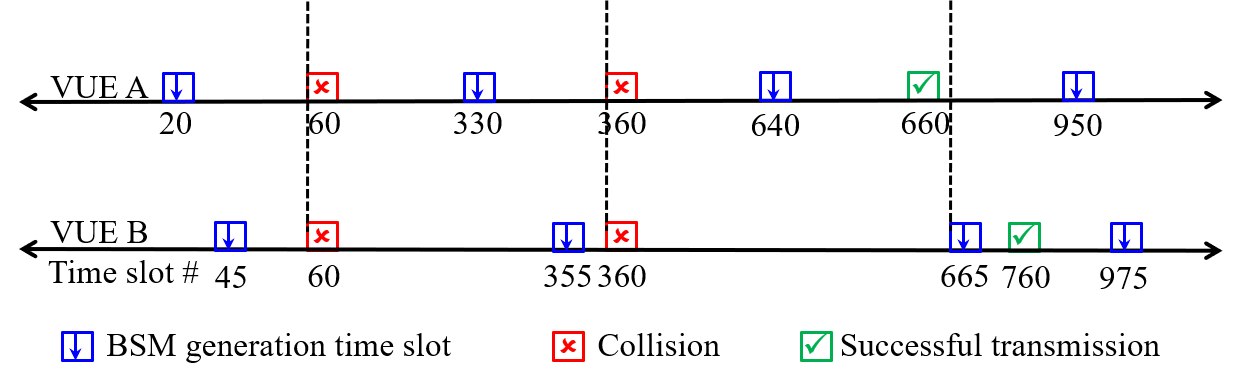}
  \caption{Example of slippage event.}\label{fig_exslippage}
  \vspace{-.25in}
  \end{center}
\end{figure}

Let VUE A and B be two interfering VUEs at a given sub-frame. Suppose that BSMs of these VUEs arrive at sub-frames $s=20$ and $45$, respectively. Also, let's assume that both VUEs have selected the same VRB in time slot $s=60$, and the congestion control is activated with $I(t)=310$ ms. Let $\mathscr{A}_{v}$ and $\mathscr{T}_v$ denote the sets of BSM generation and transmission sub-frames for the $v^{\mathrm{th}}$ VUE, respectively. Hence, we have $\mathscr{A}_\mathrm{A}=\left\{20,\,330,\,640,\,950,\dots\right\}$ and $\mathscr{T}_\mathrm{A}=\left\{60,\,360,\,660,\,960,\dots\right\}$, where each sub-frame in $\mathscr{A}_\mathrm{A}$ is increased by $I(t)$ and sub-frames in $\mathscr{T}_\mathrm{A}$ represent the $60^{\mathrm{th}}$ sub-frame of a sliding 100 ms SPS selection window. Similarly, $\mathscr{A}_\mathrm{B}=\left\{45,\,355,\,665,\,975,\dots\right\}$, and $\mathscr{T}_\mathrm{B}=\left\{60,\,360,\,760,\,1060,\dots\right\}$ for VUE B. This example is illustrated in Fig~\ref{fig_exslippage}, in which, both vehicles suffer from collisions at the first and second transmission opportunities, and go out-of-phase (i.e., VUE B slips) starting from the third transmission opportunity. Note that the sub-frame axis is not plotted to scale and we only use it for illustrative purposes. Note also that (as mentioned earlier) \textit{slippage} can only happen when $I(t)$ is not an integer multiple of 100 ms. 
 
\begin{figure}
  \begin{center}
  \includegraphics[width=8cm,height=8cm,,keepaspectratio]{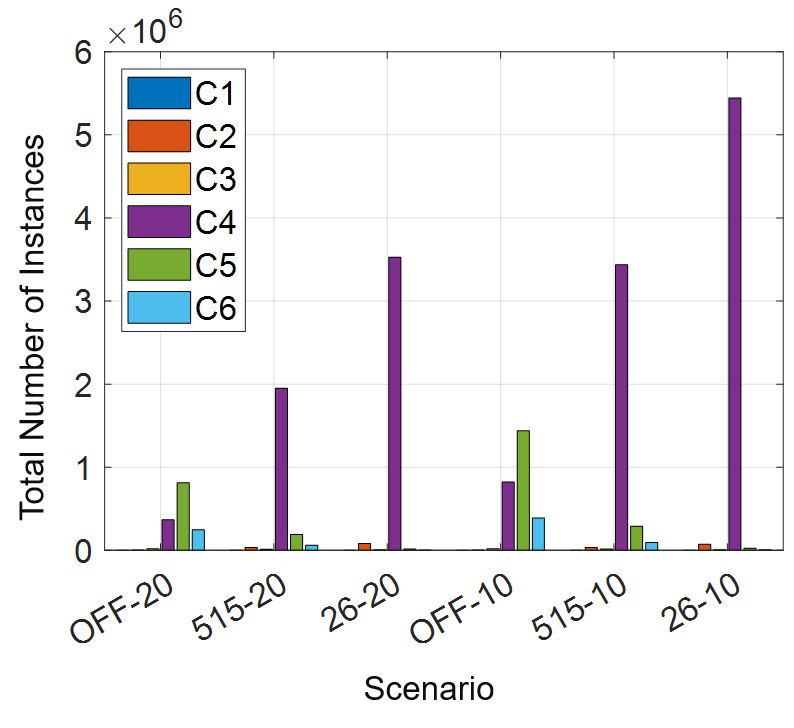}
  \caption{Classification of reasons for ending IPGs.}\label{fig_ipgend}
  \vspace{-.25in}
  \end{center}
\end{figure}

\begin{table}
\renewcommand{\arraystretch}{0.75}
  \centering
  \caption{Description of Fig.~\ref{fig_ipgend} labels.}\label{tab_ipgclass}
        \begin{tabular}{l l}
        \hline
            \textbf{Class}  & \textbf{Description}\\
        \hline
            C1 & Signal power is above thermal noise\\
        \hline
            C2  & reselection stopped IPG caused by HD transmission\\
        \hline
           C3   & Slippage stopped IPG caused by HD transmission\\
        \hline
            C4  & reselection stopped IPG caused by packet collision\\
        \hline
            C5   & Slippage stopped IPG caused by packet collision\\
        \hline
            C6   & Better conditions stopped IPG caused by deep fading\\
            \hline
        %
        \end{tabular}
        \vspace{-.25in}
\end{table}


The role of slippage is validated in Fig.~\ref{fig_ipgend} with the data from our simulations, which classifies the reasons for ending the BSM IPGs. Table~\ref{tab_ipgclass} describes the labels for this graph. Here, the data are collected at a V2V distance bin of 200 m for 400 VUE/km where $I(t) \approx$ 310 ms (see~(\ref{eq_cc}) and Table~\ref{tab_simpara}). Similar trends were observed at other V2V distances. Fig~~\ref{fig_ipgend} shows that resource reselection and slippage are the dominant reasons for ending long BSM IPGs in such cases. To better approximate these cases, we modify (\ref{eq_tk}) as:
\begin{equation}\label{eq_tk_slip}
    P(T>k)=\left(\prod^{k}_{i=1}(1-p_{\mathrm{u}}^{(i)})\right).\left(1-\sum_{i=1}^{k}q_{\mathrm{a}}^{(i)}\right),
\end{equation}
where $q_{\mathrm{a}}^{(k)}$ denotes the slippage probability conditioned on $k$ failures which can be approximated as follows:\footnote{This approximation is derived by assuming that the BSM generation times and the selected VRB timeslot are uniformly generated.} 
\begin{equation}
    q_{\mathrm{a}}^{(k)}=
        \begin{cases}
            \frac{\Psi}{\xi-T_{1}}\left(\frac{\varphi}{\xi-T_{1}}+\frac{\varphi+T_{1}}{\xi-T_{1}}\right) & \,\,\,k \le{\Psi-T_{1}},\\
            2{\Gamma}^{2}+\Gamma\left(\frac{\Psi-T_{1}}{\xi-T_{1}}\right)& \,\,\,\Psi-T_{1}<{k}\le{\Psi+T_{1}},\\
            \Gamma^{2} &\,\,\, k>{\Psi+T_{1}},
        \end{cases}
\end{equation}
where $\Psi=I(t)\mod \xi$ denotes the remainder of dividing $I(t)$ by the PDB $\xi$ and $\varphi$ is calculated as $\varphi=\left(\xi-T_{1}-k\Psi\right)$. Here, $\Gamma$ calculates the probability that either the transmitter or interferer slips after at least $\left(\Psi+T_{1}\right)$ failures as: 
\begin{equation}
    \Gamma=\frac{T_{1}}{\xi-T_{1}}.
\end{equation}

\begin{figure}
  \begin{center}
  \includegraphics[width=8cm,height=8cm,,keepaspectratio]{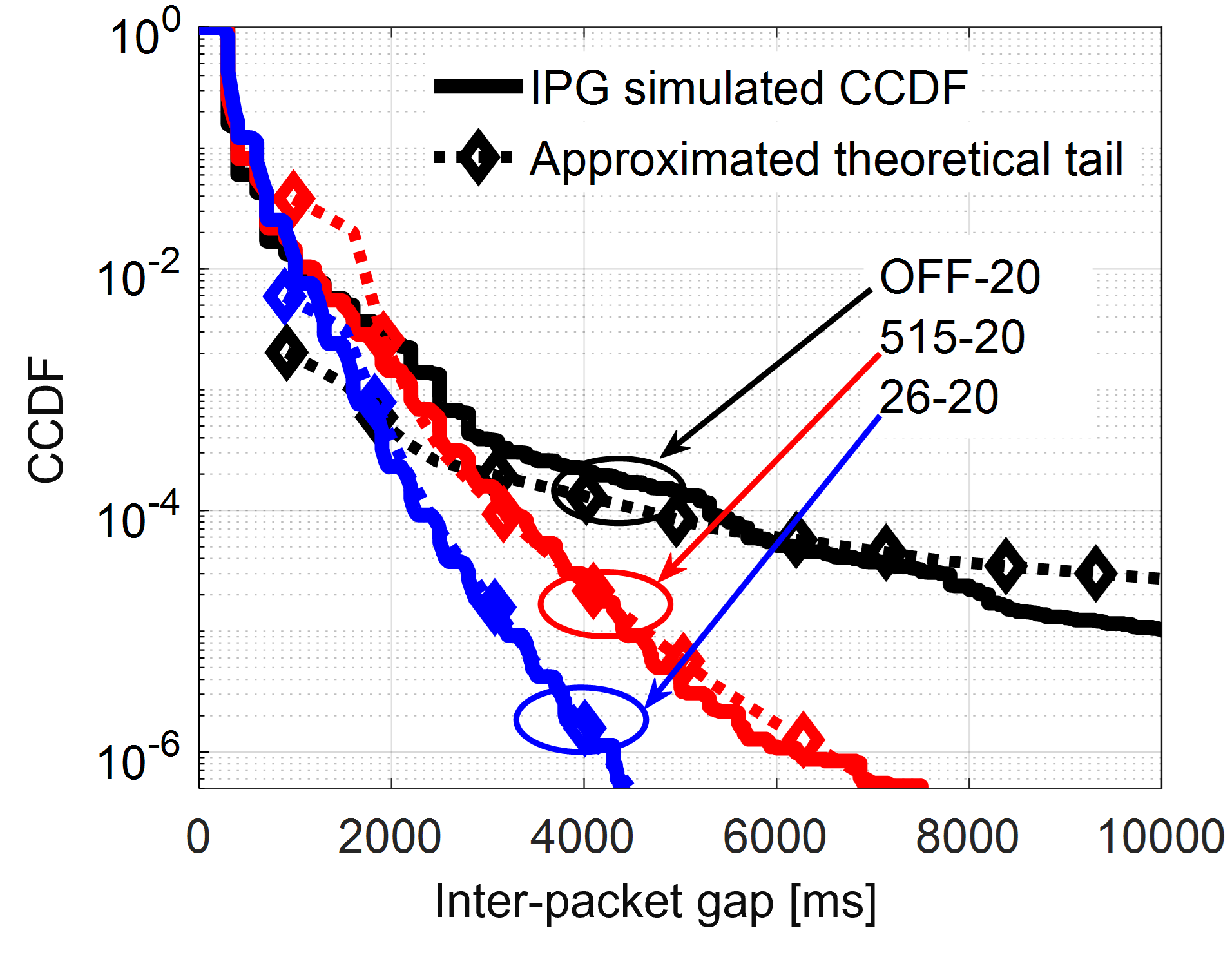}
  \caption{Approximated IPG tails: 400 VUE/km, 20 MHz}\label{fig_vald20}
  \vspace{-.25in}
  \end{center}
\end{figure}

\begin{figure}
  \begin{center}
  \includegraphics[width=8cm,height=8cm,,keepaspectratio]{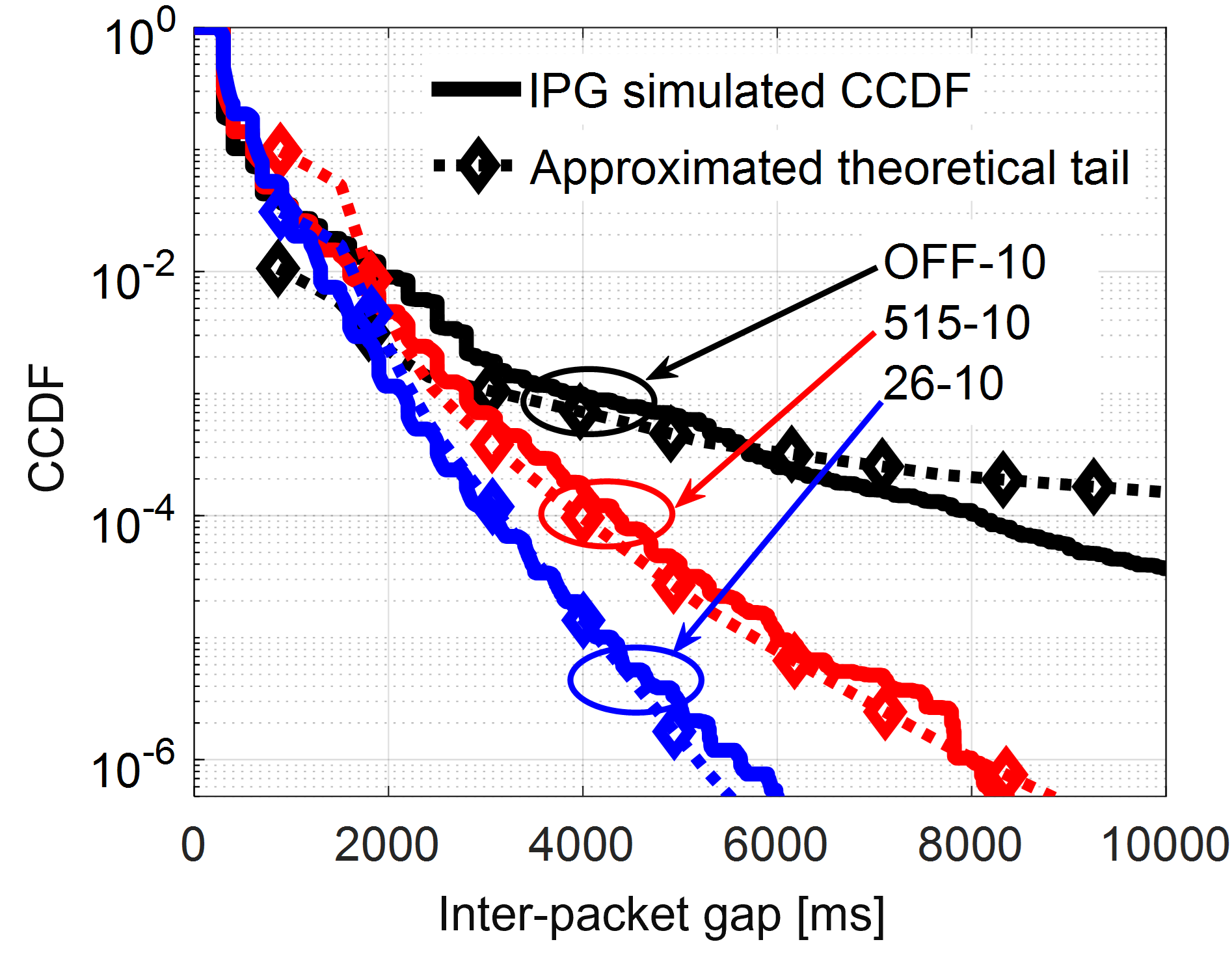}
  \caption{Approximated IPG tails: 400 VUE/km, 10 MHz}\label{fig_vald10}
  \vspace{-.25in}
  \end{center}
\end{figure}

Figs.~\ref{fig_vald20} and~\ref{fig_vald10} show the robustness of the approximated slopes accuracy when this correction term is used, again with  a vehicle density of 400 VUE/km with different one-shot and bandwidth configurations. From these figures, we see that the analytical slopes are a close match for the simulations. This suggests that again the assumptions made in deriving these approximations capture the main causes of large IPGs in the simulations. For example, without congestion control, these large IPGs are due to one dominant interferer and most likely end when the interferer or the transmitter reselects. While with congestion control, slippage in addition to SPS and one-shot reselections are both important factors.

\section{Conclusion}\label{sec_conc}
This paper investigates the impact of the SAE-standardized one-shot transmission feature to decrease the number of successive BSM losses at destination VUEs in C-V2X networks. Interleaved one-shot SPS transmissions are evaluated via extensive discrete-event simulations using a high-fidelity C++ simulator that closely follows the SPS process of C-V2X PC5 transmission mode 4. Our numerical analysis shows that using the interleaved one-shot SPS transmissions significantly improves the IPG and IA CCDF tails in different C-V2X deployment scenarios, while only slightly decreasing the PRR. Further, we develop an analytical model to estimate the approximate IPG tail behavior of BSM transmissions in C-V2X systems, which is shown to be a good match to our simulation results. This model shows that when BSM generation times are a multiple of $100$ ms, long IPG tails end due to reselection by a transmitter or a dominant interferer.  However, when the BSM generation times are not a multiple of $100$ ms, an additional {\it slippage} effect must also be accounted for.

\balance
\bibliographystyle{IEEEtran}
\bibliography{IEEEabrv.bib,Bibliography.bib}

\begin{thebibliography}{10}
\providecommand{\url}[1]{#1}
\csname url@samestyle\endcsname
\providecommand{\newblock}{\relax}
\providecommand{\bibinfo}[2]{#2}
\providecommand{\BIBentrySTDinterwordspacing}{\spaceskip=0pt\relax}
\providecommand{\BIBentryALTinterwordstretchfactor}{4}
\providecommand{\BIBentryALTinterwordspacing}{\spaceskip=\fontdimen2\font plus
\BIBentryALTinterwordstretchfactor\fontdimen3\font minus \fontdimen4\font\relax}
\providecommand{\BIBforeignlanguage}[2]{{%
\expandafter\ifx\csname l@#1\endcsname\relax
\typeout{** WARNING: IEEEtran.bst: No hyphenation pattern has been}%
\typeout{** loaded for the language `#1'. Using the pattern for}%
\typeout{** the default language instead.}%
\else
\language=\csname l@#1\endcsname
\fi
#2}}
\providecommand{\BIBdecl}{\relax}
\BIBdecl

\bibitem{vnc}
A.~Fouda, R.~Berry, and I.~Vukovic, ``Interleaved one-shot semi-persistent scheduling for {BSM} transmissions in {C-V2X} networks,'' in \emph{Proc. {IEEE} Veh. Netw. Conf. (VNC)}, Nov. 2021, pp. 143--150.

\bibitem{toyota}
T.~{Shimizu}, B.~{Cheng}, H.~{Lu}, and J.~{Kenney}, ``Comparative analysis of {DSRC} and {LTE-V2X PC5} mode 4 with {SAE} congestion control,'' in \emph{Proc. {IEEE} Veh. Netw. Conf. (VNC)}, New York, NY, USA, Dec. 2020, pp. 1--8.

\bibitem{Renault}
V.~{Mannoni}, V.~{Berg}, S.~{Sesia}, and E.~{Perraud}, ``A comparison of the {V2X} communication systems: {ITS-G5} and {C-V2X},'' in \emph{Proc. {IEEE} Vehic. Technol.Conf. (VTC-Spring)}, Kuala Lumpur, Malaysia, Jun. 2019, pp. 1--5.

\bibitem{asilomar}
A.~{Bazzi}, ``Congestion control mechanisms in {IEEE} 802.11p and sidelink {C-V2X},'' in \emph{Proc. 53rd Asilomar Conf. on Signal, Syst., Comput.}, Pacific Grove, CA, USA, Nov. 2019, pp. 1125--1130.

\bibitem{3gpp37885}
Technical{\,\,}Specification{\,\,}Group{\,\,}RAN, ``Study on evaluation methodology of new vehicle-to-everything ({V2X}) use cases for {LTE} and {NR},'' 3GPP TR37.885 v15.3.0, Jun. 2019.

\bibitem{3gpp38885}
------, ``{NR}; study on nr vehicle-to-everything {(V2X)},'' 3GPP TR 38.885 v16.0.0, Mar. 2019.

\bibitem{nrpos}
A.~Fouda, R.~Keating, and A.~Ghosh, ``Dynamic selective positioning for high-precision accuracy in {5G NR V2X} networks,'' in \emph{Proc. {IEEE} Vehic. Technol.Conf. (VTC-Spring)}, Helsinki, Finland, Apr. 2021.

\bibitem{5gaa_slr}
5G{\,\,}Automotive{\,\,}Association, ``{C-V2X} use cases volume {II}: Examples and service level requirements,'' 5GAA, Oct. 2020.

\bibitem{3gpp36213}
Technical{\,\,}Specification{\,\,}Group{\,\,}RAN, ``Evolved universal terrestrial radio {(E-UTRA)}; physical layer procedures,'' 3GPP TS 36.213 v16.4.0, Dec. 2020.

\bibitem{3gpp36321}
------, ``Evolved universal terrestrial radio {(E-UTRA)}; medium control {(MAC)} protocol specification,'' 3GPP TS 36.321 v16.3.0, Dec. 2020.

\bibitem{5gaa_PerforTest}
5G{\,\,}Automotive{\,\,}Association, ``{V2X} functional and performance test procedures – selected assessment of device to device communication aspects,'' 5GAA P-180092, Oct. 2018.

\bibitem{bspots}
A.~Bazzi, C.~Campolo, A.~Molinaro, A.~O. Berthet, B.~M. Masini, and A.~Zanella, ``On wireless blind spots in the {C-V2X} sidelink,'' \emph{{IEEE} Trans. Veh. Technol.}, vol.~69, no.~8, pp. 9239--9243, Aug. 2020.

\bibitem{J2735}
SAE{\,\,}International, ``V2x communications message set dictionary,'' J2735\_202007, Jul. 2020.

\bibitem{J3161}
------, ``On-board system requirements for {LTE-V2X V2V} safety communications,'' SAE-J3161/1, Jan. 2020.

\bibitem{tm4anal}
M.~{Gonzalez-Martín}, M.~{Sepulcre}, R.~{Molina-Masegosa}, and J.~{Gozalvez}, ``Analytical models of the performance of {C-V2X} mode 4 vehicular communications,'' \emph{{IEEE} Trans. Veh. Technol.}, vol.~68, no.~2, pp. 1155--1166, Feb. 2019.

\bibitem{piggyback}
F.~{Peng}, Z.~{Jiang}, S.~{Zhang}, and S.~{Xu}, ``Age of information optimized {MAC} in {V2X} sidelink via piggyback-based collaboration,'' \emph{{IEEE} Trans. Wireless Commun.}, vol.~20, no.~1, pp. 607--622, Jan. 2021.

\bibitem{keep}
L.~Baldesi, L.~Maccari, and R.~Lo~Cigno, ``Keep it fresh: Reducing the age of information in {V2X} networks,'' in \emph{Proc. {ACM} Int. Symp. on Mobile Ad-Hoc Netw. and Comput. (Mobihoc)}, Catania Italy, Jul. 2019, pp. 7--12.

\bibitem{AnalVNC}
B.~{Toghi}, M.~{Saifuddin}, H.~N. {Mahjoub}, M.~O. {Mughal}, Y.~P. {Fallah}, J.~{Rao}, and S.~{Das}, ``Multiple access in cellular {V2X}: Performance analysis in highly congested vehicular networks,'' in \emph{Proc. {IEEE} Veh. Netw. Conf. (VNC)}, Taipei, Taiwan, Dec. 2018, pp. 1--8.

\bibitem{TM4Config}
R.~{Molina-Masegosa}, J.~{Gozalvez}, and M.~{Sepulcre}, ``Configuration of the {C-V2X} mode 4 sidelink {PC5} interface for vehicular communication,'' in \emph{Proc. 14th Int. Conf. on Mobile Ad-Hoc and Sens. Netw. (MSN)}, Shenyang, China, Dec. 2018, pp. 43--48.

\bibitem{Apc}
M.~{Saifuddin}, M.~{Zaman}, B.~{Toghi}, Y.~P. {Fallah}, and J.~{Rao}, ``Performance analysis of cellular-{V2X} with adaptive selective power control,'' in \emph{Proc. {IEEE} 3rd Connected and Autom. Veh. Symp. (CAVS)}, Victoria, BC, Canada, Dec. 2020, pp. 1--7.

\bibitem{spatio}
B.~{Toghi}, M.~{Saifuddin}, M.~O. {Mughal}, and Y.~P. {Fallah}, ``Spatio-temporal dynamics of cellular {V2X} communication in dense vehicular networks,'' in \emph{Proc. {IEEE} 2nd Connected and Autom. Veh. Symp. (CAVS)}, Honolulu, HI, USA, Sep. 2019, pp. 1--5.

\bibitem{esps}
X.~{Wen}, M.~{Peng}, X.~{Zhang}, S.~{Yan}, and Y.~{Li}, ``Enhanced sensing-based resource scheduling algorithm for {5G V2V} communications,'' in \emph{Proc. {IEEE} Int. Conf. on Commun. in China (ICCC)}, Changchun, China, Aug. 2019, pp. 395--400.

\bibitem{AugRA}
Y.~{Jeon} and H.~{Kim}, ``An explicit reservation-augmented resource allocation scheme for {C-V2X} sidelink mode 4,'' \emph{{IEEE} Access}, vol.~8, pp. 147\,241--147\,255, Aug. 2020.

\bibitem{xuwork}
X.~Wang, R.~A. Berry, I.~Vukovic, and J.~Rao, ``A fixed-point model for semi-persistent scheduling of vehicular safety messages,'' in \emph{Proc. {IEEE} Vehic. Technol.Conf. (VTC-Fall)}, Chicago, IL, USA, Aug. 2018, pp. 1--5.

\bibitem{HIL}
G.~{Shah}, M.~{Saifuddin}, Y.~P. {Fallah}, and S.~D. {Gupta}, ``{RVE-CV2X}: A scalable emulation framework for real-time evaluation of {CV2X-based} connected vehicle applications,'' in \emph{Proc. {IEEE} Veh. Netw. Conf. (VNC)}, New York, NY, USA, Dec. 2020, pp. 1--8.

\bibitem{ProbAoI1}
M.~K. {Abdel-Aziz}, S.~{Samarakoon}, C.~{Liu}, M.~{Bennis}, and W.~{Saad}, ``Optimized age of information tail for ultra-reliable low-latency communications in vehicular networks,'' \emph{{IEEE} Trans. Commun.}, vol.~68, no.~3, pp. 1911--1924, Mar. 2020.

\bibitem{ProbAoI2}
M.~K. {Abdel-Aziz}, S.~{Samarakoon}, M.~{Bennis}, and W.~{Saad}, ``Ultra-reliable and low-latency vehicular communication: An active learning approach,'' \emph{{IEEE} Commun. Lett.}, vol.~24, no.~2, pp. 367--370, Feb. 2020.

\bibitem{RRM}
X.~{Chen}, C.~{Wu}, T.~{Chen}, H.~{Zhang}, Z.~{Liu}, Y.~{Zhang}, and M.~{Bennis}, ``Age of information aware radio resource management in vehicular networks: A proactive deep reinforcement learning perspective,'' \emph{{IEEE} Trans. Wireless Commun.}, vol.~19, no.~4, pp. 2268--2281, Apr. 2020.

\bibitem{3gpp36212}
Technical{\,\,}Specification{\,\,}Group{\,\,}RAN, ``Evolved universal terrestrial radio {(E-UTRA)}; multiplexing and channel coding,'' 3GPP TS 36.212 v16.4.0, Dec. 2020.

\bibitem{3gpp36331}
------, ``Evolved universal terrestrial radio {(E-UTRA)}; radio resource control {(RRC)}; protocol specification,'' 3GPP TS 36.331 v16.3.0, Dec. 2020.

\bibitem{J2945}
SAE{\,\,}International, ``On-board system requirements for {V2V} safety communications,'' SAE-J2945/1, Mar. 2016.

\bibitem{capacity}
P.~{Gupta} and P.~R. {Kumar}, ``The capacity of wireless networks,'' \emph{{IEEE} Trans. Inf. Theory}, vol.~46, no.~2, pp. 388--404, Mar. 2000.

\bibitem{ITUpathloss}
International{\,\,}Telecommunications{\,\,}Union, ``Propagation data and prediction methods for the planning of short-range outdoor radio communication systems and radio local area networks in the frequency range 300 {MHz} to 100 {GHz},'' Geneva, Switzerland, ITU-R P.1411-10, Aug. 2019.

\bibitem{SimulationModeling}
A.~M. Law, \emph{Simulation Modeling and Analysis}.\hskip 1em plus 0.5em minus 0.4em\relax McGraw-Hill Education, 2015.

\bibitem{D.Tse}
D.~Tse and P.~Viswanath, \emph{Fundamentals of Wireless Communication}.\hskip 1em plus 0.5em minus 0.4em\relax Cambridge Univ. Press, 2005.

\bibitem{NIST}
J.~Wang and R.~Rouil, ``{BLER} performance evaluation of {LTE} device-to-device communications,'' NISTIR 8157, Nov. 2016.

\bibitem{3gpp36214}
Technical{\,\,}Specification{\,\,}Group{\,\,}RAN, ``Evolved universal terrestrial radio {(E-UTRA)}; physical layer; measurements,'' 3GPP TS36.214 v16.1.0, Jun. 2020.

\bibitem{etsi}
European{\,\,}Telecommunications{\,\,}Standards{\,\,}Institute, ``Intelligent transport systems {ITS}; congestion control mechanisms for the {C-V2X PC5} interface; layer part,'' ETSI TS 103 574 V1.1.1, Nov. 2018.

\end{thebibliography}
\end{document}